\documentclass[12pt]{iopart}
\usepackage{graphicx}
\usepackage{epstopdf}
\usepackage{subfigure}
\usepackage[authoryear]{natbib}
\usepackage{longtable}
\usepackage{bm}
\usepackage{url}
\usepackage{caption}
\usepackage{microtype}
\begin{document}

\title{An inevitably aging world - Analysis on the evolutionary pattern of age structure in 200 countries}

\author{Jiajun Ma$^1$, Qinghua Chen$^{1*}$, Xiaosong Chen$^{1}$, Jingfang Fan$^{1}$, Xiaomeng Li$^{1*}$, Yi Shi$^2$}

\address{1. School of Systems Science, Beijing Normal University, Beijing, China}
\address{2. China Population and Development Research Center, Beijing, China}

\ead{lixiaomeng@bnu.edu.cn, qinghuachen@bnu.edu.cn}
\vspace{10pt}

\begin{abstract}

Ignoring the differences between countries, human reproductive and dispersal behaviors can be described by some standardized models, so whether there is a universal law of population growth hidden in the abundant and unstructured data from various countries remains unclear. The age-specific population data constitute a three-dimensional tensor containing more comprehensive information. The existing literature often describes the characteristics of global or regional population evolution by subregion aggregation and statistical analysis, which makes it challenging to identify the underlying rules by ignoring national or structural details. Statistical physics can be used to summarize the macro characteristics and evolution laws of complex systems based on the attributes and motions of masses of individuals by decomposing high-dimensional tensors. Specifically, it can be used to assess the evolution of age structure in various countries over the past approximately 70 years, rather than simply focusing on the regions where aging has become apparent. It provides a universal scheme for the growing elderly and working age populations, indicating that the demographics on all continents are inevitably moving towards an aging population, including the current ``young'' continents of Africa, and Asia, South America with a recent ``demographic dividend''. It is a force derived from the ``life cycle'', and most countries have been unable to avoid this universal evolutionary path in the foreseeable future.


\end{abstract}

\maketitle

\section{Introduction}

After World War II, ``peace and development'' became the main topic for most countries/regions. In addition, the world population expanded from 2.5 billion in 1950 to 8 billion in 2022. The United Nations predicts it could continue to grow to approximately 8.5 billion in 2030, 9.7 billion in 2050 and 10.4 billion in 2100~\citep{united2022world}. Humans play the dual role of producer/creator and  consumer/destroyer of resources. The rapid growth of the population and the vigor of urbanization have increasingly put pressure on the natural environment and resources~\citep{khan2022links}. On the other hand, as a key component of production and the main avenue for society and culture, population is an indispensable foundation for a country and the inheritance of culture. Therefore, Science/AAA's ``125 Questions: exploration and discovery'' wonders: will the world population keep growing indefinitely?~\citep{sanders2021125}. Related studies on fertility, mortality, migration and aging of the population have always been the concern of scholars and governments~\citep{bloom2015global,liu2020education,gu2021major} and the basis for research in many fields, such as economics, trade, environmentalism, and politics~\citep{woods2007ancient}.

Age structure determines the potential for future growth of a particular age group, as well as the overall population, and populations in different age groups have various effects on economic production and social consumption, so the study of population age structure cannot be ignored. In the 21st century, there have been two major underlying demographic shifts, aging and urbanization, which are also drivers of significant social transformation~\citep{beard2012ageing,bloom2016global}. Therefore, in addition to the total population, heterogeneity and the evolution of the age structure are also important demographic issues~\citep{harper2014economic,ritchie2019age,bai2020new}. The age-specific population data constitute a complex tensor with three dimensions: time, country and age group. Due to the lack of appropriate systematic analysis methods, scholars always sum, average or subtract data in one or two dimensions. For example, some publications describe the total population, growth rate, and age structure, including the time evolution and spatial distribution characteristics of the world or specific regions, with the loss of national or structural details~\citep{permanyer2019global,gu2021major}. Other studies focused on specific countries to analyze their structural, temporal and evolution characteristics~\citep{liu2010china,angel2017aging}. Even if data from each country could be analyzed separately, this would not help uncover the universal rules of human society, nor enable us to describe and forecast the future age structure.

In previous studies, scholars found that, even in different geographical locations, economic levels and cultural backgrounds, human life cycles and age-specific mortality always follow similar laws~\citep{wang2017global,aburto2020dynamics}. Therefore, it is natural to conclude that humans have similar reproductive and migration selection behaviors~\citep{aksoy2021refugees}, and relevant models have been proposed and have achieved good fitting results~\citep{li2016characterizing,huang2020modelling,huang2021longcuts}. Based on the similarity of human choice behavior, we wonder whether there is a universal law of population growth hidden in the abundant and unstructured data from various countries. In the socioeconomic system, we can also extract some rules that are less affected by differences in economies, societies and cultures, similar to the rules in most natural systems. Some researchers have proposed the objective paths of product structure upgrading and economic complexity growth based on high-dimensional international trade data~\citep{Hidalgo2007The,2018nature,2021Economic}. Similarly, ``Is aging an inevitable trend that no country can avoid, even if some countries take related measures?'' To answer this question, we should not only focus on the regions where these phenomena have occurred, but also explore the basic path of age structure evolution from high-dimensional data.

Eigen microstate theory originates from statistical physics, which is used to analyze the macroscopic behaviors of a system composed of multiple interacting objects~\citep{li2016critical,hu2019condensation} and has played a role in the analysis of different complex systems in nature and economic society~\citep{sun2021eigen,liu2022renormalization}. Unlike traditional statistical analysis, the eigen microstate method does not reduce the dimension of high-dimensional tensors, but explores some macro phenomena and evolution laws of the system and the collective behaviors of objects by integrating information from different dimensions~\citep{li2021discontinuous}. Examples include the prediction of El Niño and La Niña events from temperature changes at 18,048 observation stations and the description of the coevolution mechanism of the energy or material sector from price fluctuation of 1600 stocks~\citep{sun2021eigen}. This method can also be used to analyze the evolution of complex systems described by age-specific population data from 200 countries.

The remainder of the paper is structured as follows: Section 2 describes the database and methods used to analyze the tensor and achieve higher-order decomposition, using less information to describe the original empirical data by identifying the macroscopic behavior, evolution rules of the entire system and the difference in individuals compliance. Section 3 describes a universal trend in changes in the elderly and working age populations, indicating that the demographics on all continents seem to be inevitably shifting towards an aging population, including the current ``young'' continents of Africa, and Asia, South America with ``demographic dividend''. Section 4 provides the conclusions and discussion.

\section{Data $\&$ Methods}

\subsection{Data Source}

The world population data were obtained from the United Nations Department of Economic and Social Affairs, including overall and age-specific population. The data covered 200 countries/regions\footnote{In this paper, ``country'' and ``countries'' are used to replace all ``country/region'' and ``countries/regions''. This is only a simplification of the text and does not represent the authors' view on whether different entities are sovereign states.} and was divided into 21 age groups. These annual population data were divided into two parts: the data from 1950 to 2021 were estimated by the United Nations, and the data from 2022 to 2050 are predicted values.

To show the differences in the evolution of age structure among countries with different income levels, we divided 200 countries into four categories according to the standards of the World Bank, including 28 low-income, 54 lower-middle-income, 48 upper-middle-income and 63 high-income countries, in addition to 7 countries or territories where the economic situation is unknown.

\begin{table}[htbp]
\tiny
  \centering
  \caption{Data description}
    \begin{tabular}{p{11em}p{27em}p{15em}}
    \hline
    Variable & Variable Description & Data Source \\
    \hline
    Population by age group & Quinquennial Population by One-Year Age Groups - Both Sexes. De facto population as of 1 July of the year indicated classified by one-year age groups (0-4, 5-9, 10-14, ..., 95-99, 100+). Data are presented in thousands. Total population (both sexes combined) by five-year age group, region, subregion and country, annually for 1950-2021 (thousands) Estimates, annually for 1950-2021 (thousands) Medium fertility variant. The list of countries and continents in Appendix A.5.
    
 &
 \url{https://population.un.org/wpp/Download/Standard/Population/.} \\
\hline
    Country economic classification & Data were from the World Bank, which divides countries into four categories according to their economic level: low-income economies, lower-middle-income economies, upper-middle-income economies, high-income economies. &    \url{https://datahelpdesk.worldbank.org/knowledgebase/articles/906519-world-bank-country-and-lending-groups} \\
    \hline
    Map template (.shp)  & The map template is from WORLD Country Polygons in the World Bank. & \url{https://datacatalog.worldbank.org/search/dataset/0038272} 
    \\
    \hline
    \multicolumn{2}{c}{Note: America etc mentioned in the passage refers to Oceania, North America and South America.} 
    \end{tabular}%
  \label{tab:tab1}%
\end{table}%

\subsection{Tensor decomposition}

The global population is a complex system that evolves with time and is composed of 200 countries, which are the agents of the system. Using the age groups $i = 1, 2, ... ,M$ and the times $t = 1, 2, ... ,L$ in sequence, we obtain the state series $S_{n}(i,t)$ of agent $n$, with $n=1,2,...,N$. In the literature, data analysis using eigen microstate theory usually focuses on two-dimensional matrices~\citep{sun2021eigen,li2021discontinuous}. To describe the system state more comprehensively, we extended the matrix data to tensors and used canonical polyadic (CP) decomposition~\citep{kolda2009tensor,anandkumar2014tensor} to obtain vectors describing different dimensions. Canonical polyadic (CP) decomposition is a method of decomposing a tensor into a sum of rank-1 tensors.

\begin{figure}
    \centering
    \includegraphics[width=0.8\linewidth]{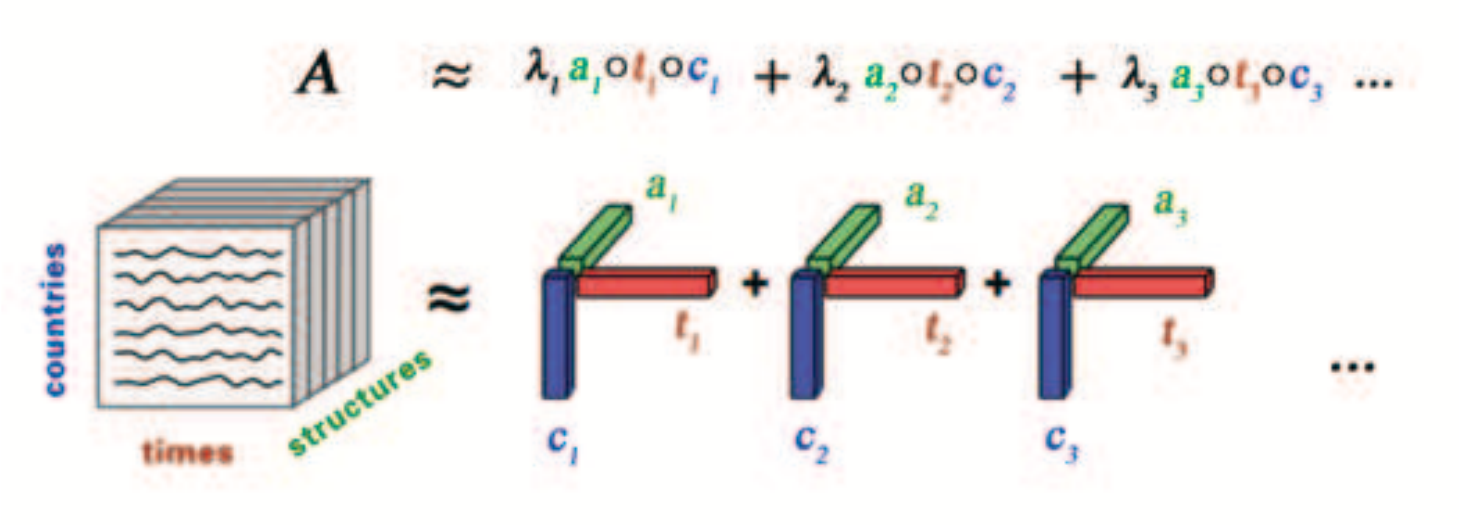}
    \caption{Canonical Polyadic (CP) Decomposition}
    \label{fig:pic1}
\end{figure}

Here, we have $M\times L \times N$ tensor $\boldsymbol{S}$, and $S_{n}(:,t)$ represents the age structure proportion of country $n$ in year $t$. Then, normalize tensor $\boldsymbol{S}$ from equation $\boldsymbol{A} = \boldsymbol{S}/norm(\boldsymbol{S})$, where $norm(\boldsymbol{S})=\sqrt{\sum_{i,t,n}S_n(i,t)^2}$. As shown in Figure \ref{fig:pic1}, according to the CP decomposition and $r=1,2,3...,R$, the tensor $\boldsymbol{A}$ can be factorized as

\begin{equation}
    \boldsymbol{A} \approx \sum\limits_{r = 1}^R {{\lambda _r}{\bm{a_r}} \circ \bm{{t_r}}}  \circ \bm{{c_r}}\\
    A_{itn}\approx\sum_{r=1}^R\lambda _r a_r(i)t_r(t)c_r(n).
    \label{eq:A}
\end{equation}

$\bm{a_r}\circ \bm{c_r}$ constitutes the $r$-th eigen microstate (EMr) of the $N$ agent system. $\bm{t_r}$ describes the evolution of EMr over time. Figure \ref{fig:pic1} shows that it reproduces complicated empirical data with less information and fewer variables. $\bm{a_r}$ represents the eigenvectors in the age structure dimension, $\bm{t_r}$ represents eigenvectors in the time dimension, and $\bm{c_r}$ represents the eigenvectors in the country dimension. Then, the description of the tensor ($M\times L \times N$) is decomposed into two parts: one is the macroscopic behavior of the whole system as $\bm{a_r}$ ($M\times R$) and $\bm{t_r}$ ($L\times R$), and the other is the quantification of each country ($N \times R$). The principle of dimensionality reduction is to use less information to describe the original empirical data by identifying the macroscopic behavior, evolution rules of the entire system and the difference in each agent's compliance. Therefore, eigen microstate theory can help us find the basic evolution patterns of the overall population and structure in the world from complex population data, as well as the personalized characteristics of different countries.

Here, we used the alternating least squares (ALS) algorithm to realize CP decomposition. After choosing the rank, we randomly initialized 500 times to avoid the local extremum and select the best-fitting solution described by the fit value. Then, we obtained the unique solution for a given rank. The fit value $f=1-norm(\boldsymbol{A}-\hat{\boldsymbol{A}})$ was loosely the proportion of the data described by the CP model, where $\bm{\hat{A}}$ is an approximate tensor. We used the Tensor toolbox (Version 3.4) of MATLAB to obtain CP decomposition results. We tested the robustness by comparing the results of 10 experiments, including CP decomposition, Singular Value decomposition (SVD) and High Order Singular Value decomposition  (HOSVD)~\citep{zeng2020decompositions,kolda2009tensor}.

\subsection{Age structure regression and coefficient analysis}

The results of the previous decomposition describe the macromorphology of the world population system, with each eigenstate describing a characteristic of the system. When analyzing the evolution pattern of different countries, the consistency of each country with these macro features and its evolution path should be quantified. Here, we standardize the age structure proportion $S_{n}(:,t)$ and describe it with $R$ characteristics as $\bm{a_1}$, $\bm{a_2}$, $\bm{a_3}$,..., $\bm{a_R}$ (the eigenvectors in the age structure dimension got by eq. \ref{eq:A}),

\begin{equation}
\label{eq:regress}
    S_{n}(:,t)=A_{n,t}*\bm{a_1}+B_{n,t}*\bm{a_2}+C_{n,t}*\bm{a_3}
\end{equation}

Parameter $A_{n,t}$ and $B_{n,t}$ and $C_{n,t}$ describe the impact of characteristics $\bm{a_n}$ on the age structure of country $n$ in year $t$. $D_{n,t}$ is the constant of regression.

\section{Results}

Based on empirical data, the tensor $\boldsymbol{A}$ describes the population evolution patterns of 200 countries by age group from 1950 to 2021. This $M\times L \times N$ tensor is difficult to analyze quantitatively from a global and dynamic perspective by traditional methods. Here, we decompose the tensor $\boldsymbol{A}$ into three dimensions: age structure, time, and country. First, we needed to choose the appropriate rank of decomposition. Determining the rank of a tensor is NP-hard~\citep{kolda2009tensor}.  Here, we used the fit value to determine the rank of the CP model (as $R=3$). Second, we performed 10 experiments. Each experiment was initialized randomly 500 times, and the final results were selected based on the fitted values and orthogonality. The CP decomposition proposes the three largest eigenmicrostates, explaining 88.65\% of the information in the data. As mentioned before, the CP decomposition can extract the evolution rules of the global population system as $\bm{a_r}$, $\bm{t_r}$, and the characteristics of each country as $\bm{c_r}$ (with $r=1,2,3$). The specific analysis is described in the subsequent sections. In addition, we used other methods, such as SVD decomposition and HOSVD decomposition, and the phenomena highlighted in other decomposition results are similar. The decomposition results and details are presented in Appendix A.1.

\subsection{Three main characteristics of global population evolution}

\subsubsection{Continuous growth was accompanied by slowing growth.}

\begin{figure}[h]
    \centering
    \subfigure[First eigen microstate as EM1]{
    \label{fig:fig1a} 
    \includegraphics[width=1\textwidth]{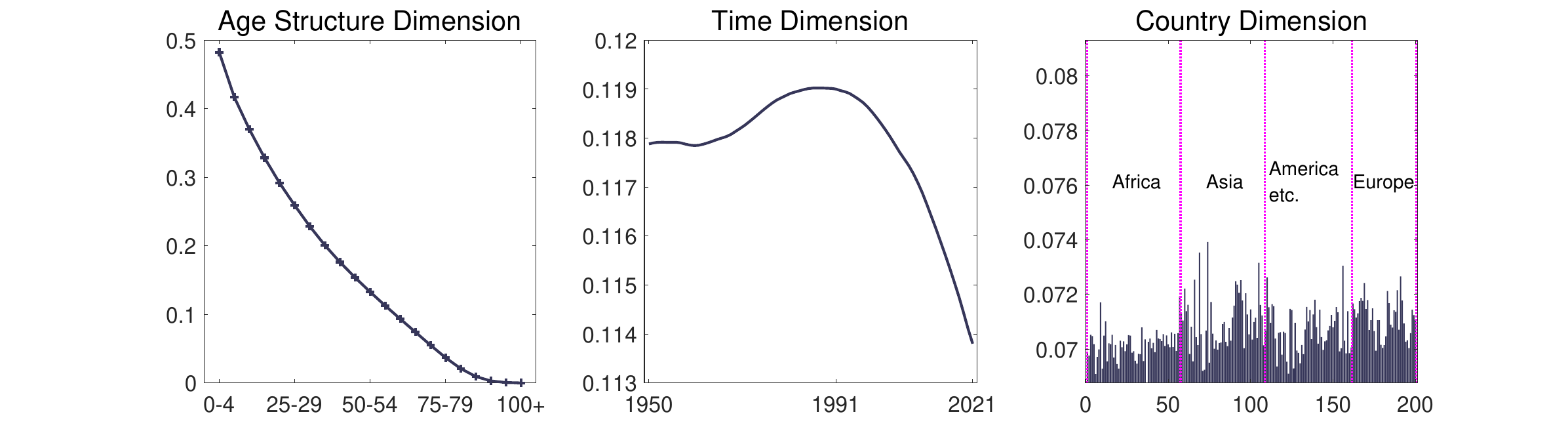}}
    \subfigure[Global population structure and comprehensive indicators]{
    \label{fig:fig1b} 
    \includegraphics[width=0.5\textwidth]{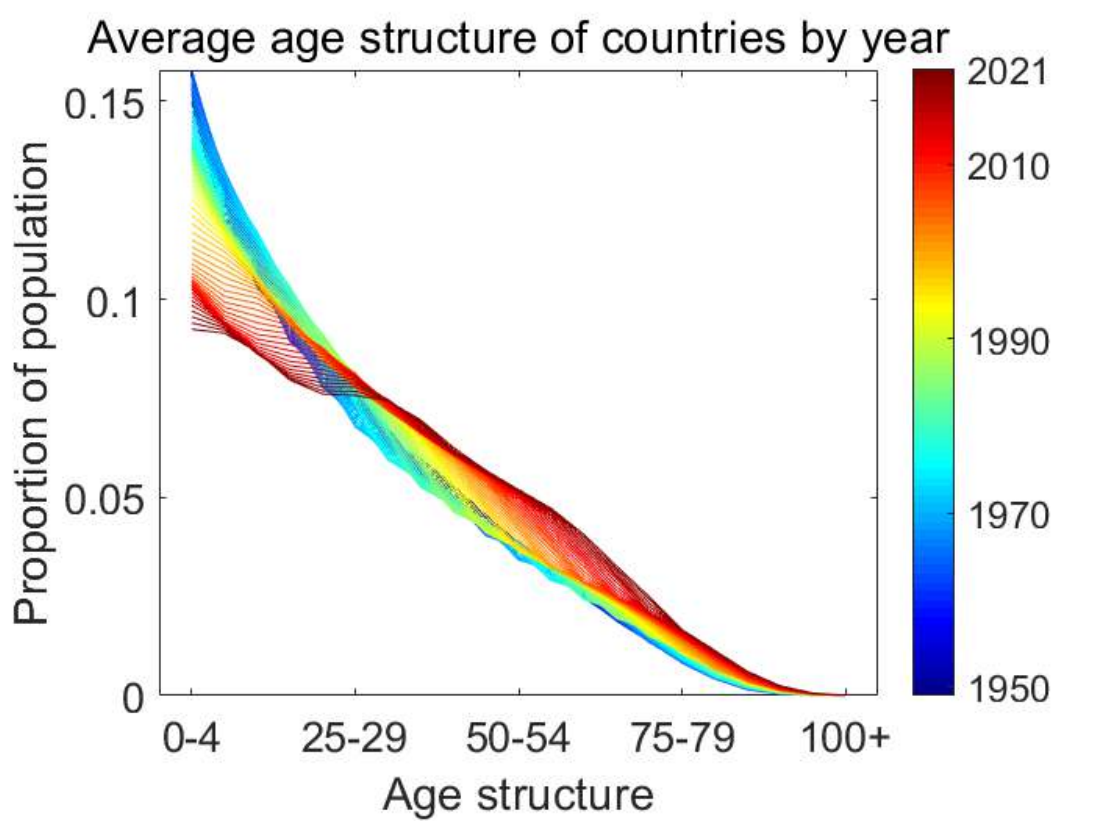}}
    \caption{Continuous population growth and gradual slowdown in various countries.}
    \label{fig:pic2}
\end{figure}

The first eigen microstate had positive $\bm{a_1}$, $\bm{t_1}$ and $\bm{c_1}$ (Figure \ref{fig:fig1a}). $\bm{c_1}$ for all countries was located in [0.0688, 0.0739], its variance was $7.9121e^{-07}$, and the gap between countries was small, indicating that the first eigenstate is universal to all countries in the world for more than 70 years after World War II (Figure \ref{fig:fig1a}, right).

For the age structure, $\bm{a_1}$ has typical pyramid features with the population proportion decreasing with increasing age. The downward convex curve is generally considered to have a population growth pattern of high fertility and low aging (Figure \ref{fig:fig1a}, left). The value range of $\bm{t_1}$ was [0.1138, 0.1730]. This growth type of age structure as $\bm{a_1}$ was generally present for 72 years, and its influence increased first and then decreased, with $\bm{t_1}$ reaching its maximum value in approximately 1991 (Figure \ref{fig:fig1a}, middle). Figure \ref{fig:fig1b} describes changes in the average age structure of all countries, from the 1950s in blue, accompanied by declining fertility rates, to the 1990s in green. It shows little change in age structure before the 1990s and then deviates rapidly from the age structure described by $\bm{a_1}$, showing a relative aging prominence and reduction in fertility (in red). It represents the trend of maintaining population growth in most countries, but the growth is gradually slowing down.

\subsubsection{Growth and coexistence of the working population and the elderly population.}

\begin{figure}[h]
    \centering
    \subfigure[Second eigen microstate as EM2]{
    \label{fig:figure2a} 
    \includegraphics[width=1\textwidth]{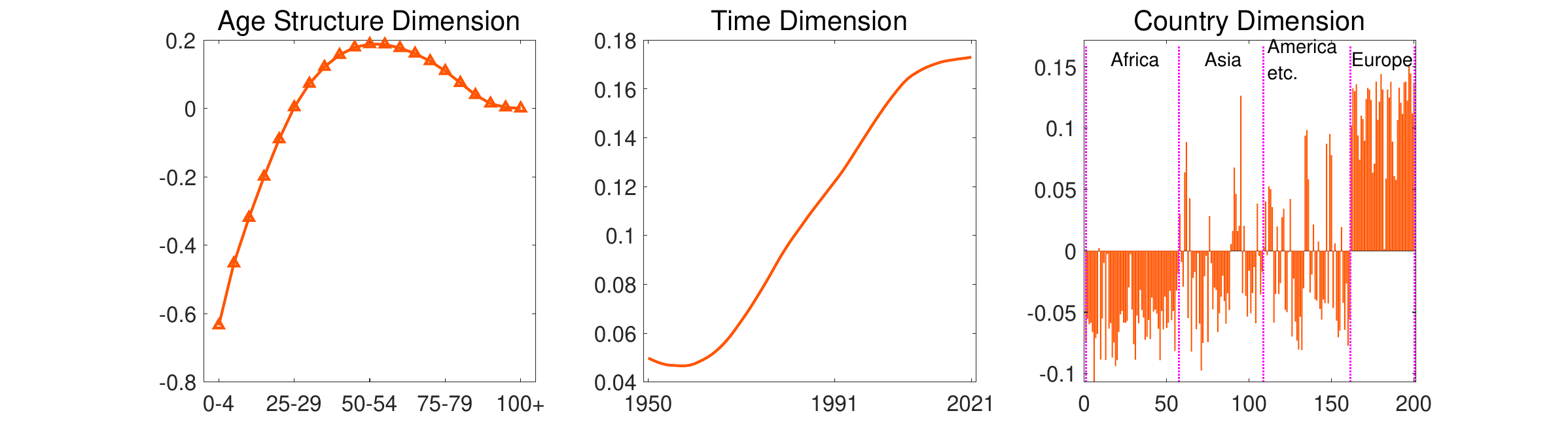}}
    \subfigure[Third eigen microstate as EM3]{
    \label{fig:figure2b} 
    \includegraphics[width=1\textwidth]{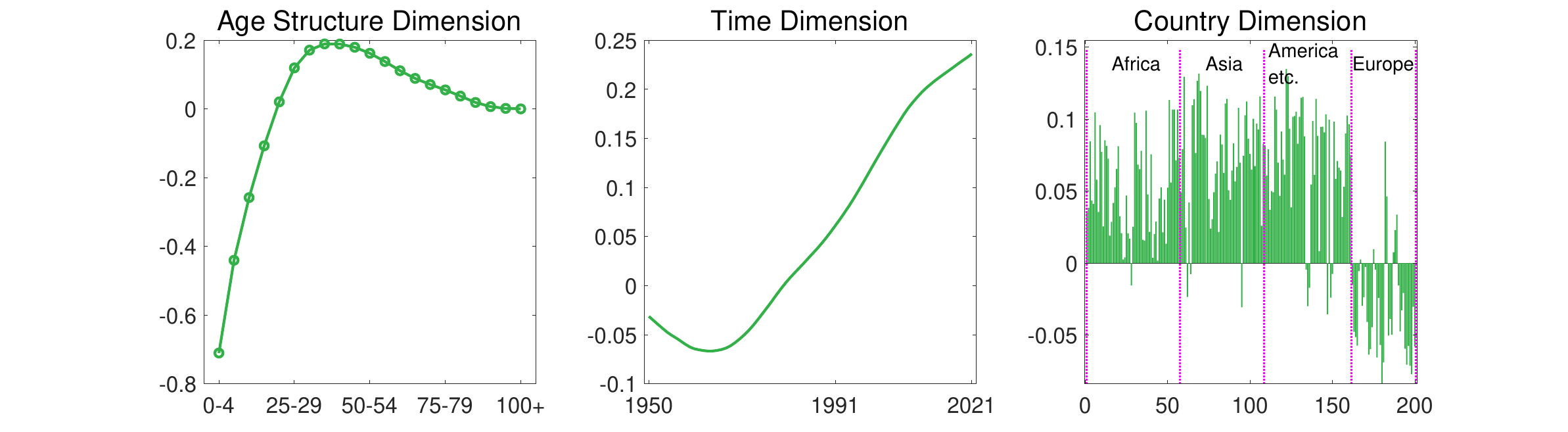}}
    \caption{Evolutionary characteristics of the elderly and labor increasing}
    \label{fig:pic33}
\end{figure}

EM2 and EM3 describe two types of evolutionary trends (Figure \ref{fig:pic33}). For the countries, $\bm{c_2}$ and $\bm{c_3}$ had both positive and negative values. Most countries had obvious values of $\bm{c_2}$ and $\bm{c_3}$, meaning that these two trends together characterize the evolution of almost all countries. The countries corresponding to the positive value of $\bm{c_2}$ are mainly concentrated in Europe, and the countries corresponding to the positive value of $\bm{c_3}$ are mainly concentrated in Africa, Asia and other regions. We draw the specific values in the map in the appendix A.2. The age structure showed some commonality of $\bm{a_2}$ and $\bm{a_3}$, that is, the phenomenon of declining birth rate and an insufficient number of teenagers. However, the specific shapes of the two age structures were different. For EM2, in terms of proportion of the population, the largest age group aged 50 to 54, and the older population structure generally showed a significant increase (that is, thick-tailed characteristics), so it could be considered to describe the obvious growth of the elderly population (Figure \ref{fig:pic33}a). In contrast, for EM3, the largest age group aged 35 to 39. Moreover, it was no longer evident before the age of 20 and after the age of 60; thus, we believe that EM3 describes the growth of the labor force structure (Figure \ref{fig:pic33}b).

For the evolutionary behavior, although $\bm{t_2}$ and $\bm{t_3}$ both exhibited a trend of decreasing at first and then increasing, $\bm{t_2}$ was always greater than zero, and $\bm{t_3}$ fell below zero between 1950 and 1979. During this time period, there could be a transient opposite evolutionary trend, such as sufficient numbers of newborns and adolescents. By comparing the slope of the curve, we found that the enhancement rate of EM2 was relatively slower than that of EM3. The two phenomena of increases in the elderly and working age populations will be analyzed in detail in the following sections.

\subsection{The evolutionary characteristics of the elderly and working age population increases}

\subsubsection{Countries change from similarity to polarization.}

For country $n$, we regressed the standardized age structure proportions in year $t$ with eq. \ref{eq:regress}, where $B_{n,t}$ and $C_{n, t}$ describe the coefficient of the elderly and labor increasing. The range of the regression result determination coefficient $R^2$ was [0.6669,1), with a mean value of 0.9783. In addition, 99.59\% of $R^2$ values were greater than 0.80, 97.77\% was greater than 0.90, and 88.28\% was greater than 0.95. Here, we describe the age structure of countries at different times using three features: $\bm{a_1}$, $\bm{a_2}$ and $\bm{a_3}$.

\begin{figure*}[h]
\centering
\subfigure[Boxplot]{\label{fig:fig55a}
\includegraphics[width=0.45\linewidth]{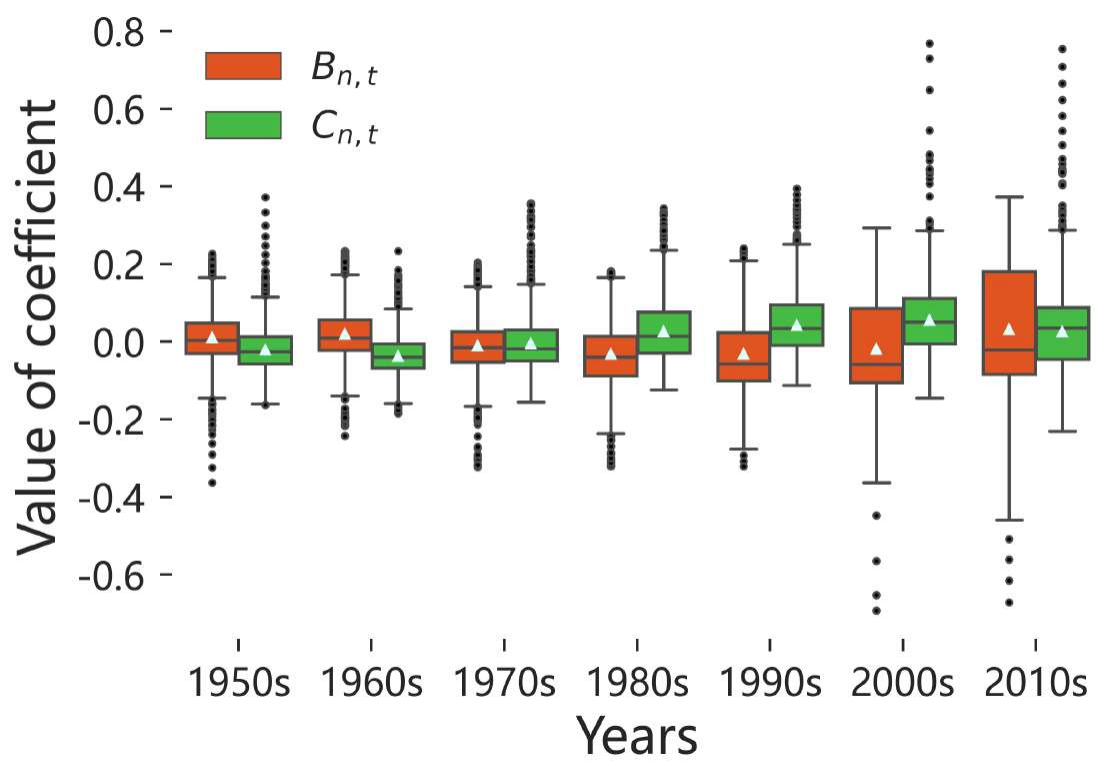}}
\hspace{0.01\linewidth}
\subfigure[Ridgeplot]{\label{fig:fig55b}
\includegraphics[width=0.5\linewidth]{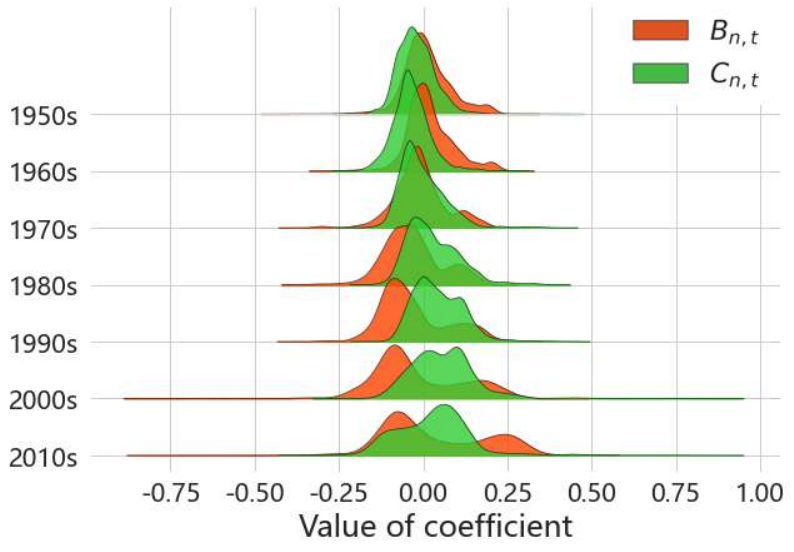}}
\caption{Distribution of regression coefficients}
\label{fig:fig5}
\end{figure*}

In Figure \ref{fig:fig55a}, the red boxes represent the distribution of $B_{n,t}$, and the green boxes represent the distribution of $C_{n,t}$ in year $t$, which describe the influence of elderly and working age population growth on the age structure in different countries. In each box, the horizontal line from bottom to top represents the minimum, the first quartile, the third quartile and the maximum value, and the white triangle represents the mean value. First, the variances of $B_{n,t}$ and $C_{n,t}$ exhibited a gradually increasing trend, indicating that the differences in population structure between countries continued to grow, and the prominence of the elderly and working age populations differed among countries. Second, the mean of $B_{n,t}$ was initially greater than $C_{n,t}$, and since the 1980s, the mean of $C_{n,t}$ has exceeded $B_{n,t}$, indicating that the change in labor structure in many countries during this period was more significant, exceeding the trend of aging. However, in the last decade, there have been many large positive values for $B_{n,t}$, meaning that the structural characteristics of elderly individuals in many countries have been more significant.

The ridgeplot shows the distribution and evolution of $B_{n,t}$ and $C_{n,t}$ in detail (Figure \ref{fig:fig55b}). The distribution of $B_{n,t}$ and $C_{n,t}$ gradually evolved from a unimodal to a bimodal distribution, reflecting the trend of countries' demographic characteristics from similarity to polarized, and the polarization of aging has become more obvious in recent years. For $C_{n,t}$, in the early years, the only peak was located in the area less than 0. In the 1980s, a positive subpeak appeared, and then it gradually increased and replaced the original negative peak to form a new maximum peak in the 2010s. The evolution of $B_{n,t}$ from a unimodal to a bimodal distribution was more obvious. Although the current positive subpeak did not exceed the negative subpeak, it continued to increase and shift to the right side, indicating that the number of countries with an aging trend is increasing, and the influence of the elderly is also strengthening. In addition, we plotted the ridgeplot from the 2020s to 2040s (Appendix A.2), and it shows that the peak with positive values overtakes the peak with negative values (Figure \ref{fig:pic88}).

\subsubsection{A universal alternating path of the elderly and working age population.}

\begin{figure*}[htp]
\centering
\subfigure[In rectangular coordinates]{\label{fig:fig51a}
\includegraphics[width=0.45\linewidth]{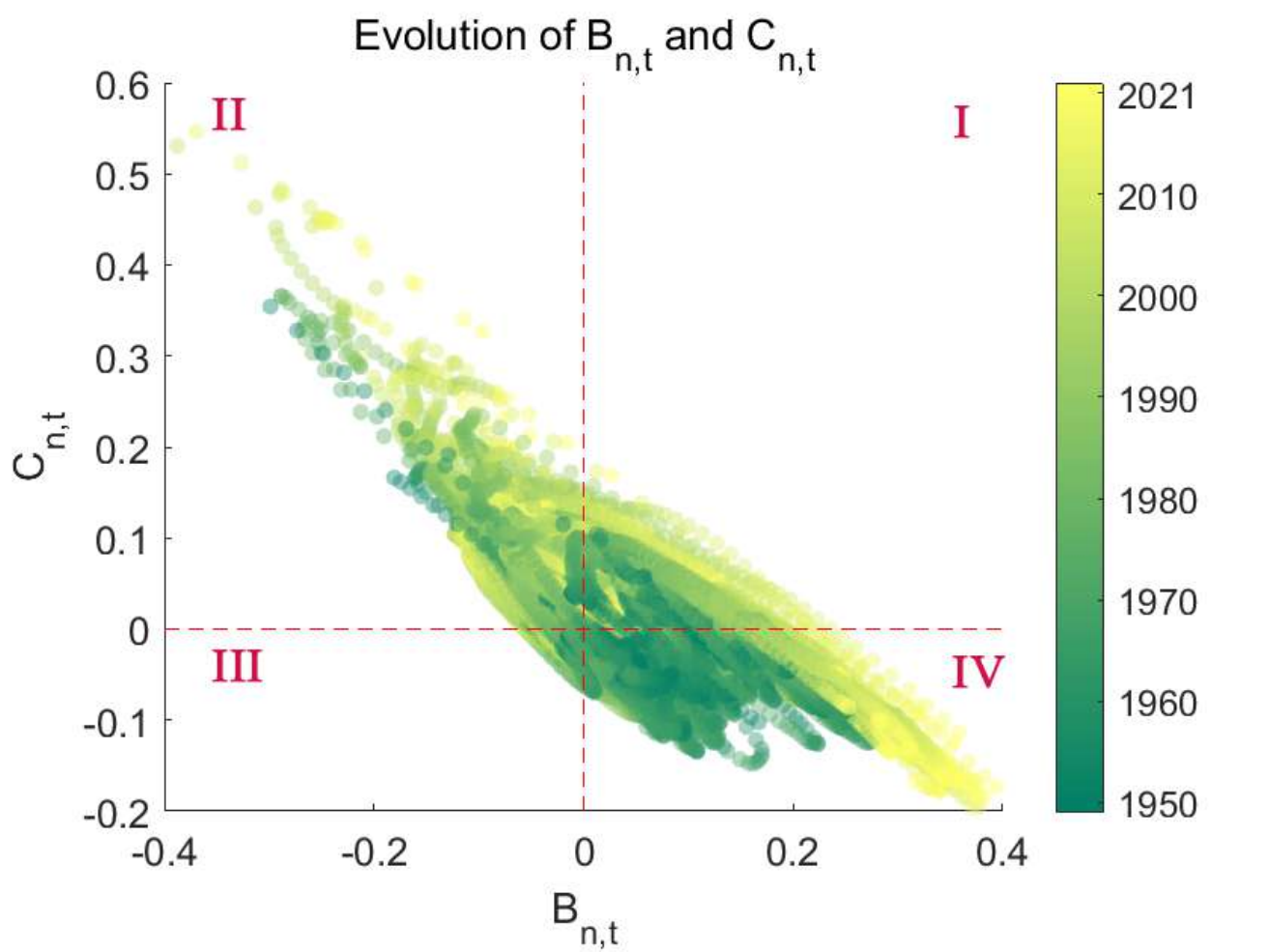}}
\hspace{0.01\linewidth}
\subfigure[In polar coordinates]{\label{fig:fig51b}
\includegraphics[width=0.45\linewidth]{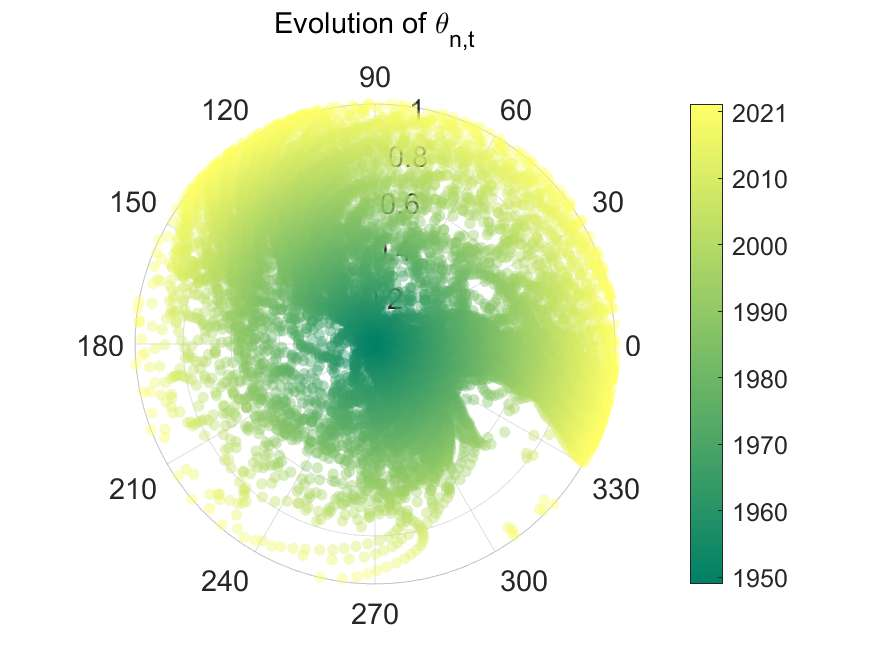}}
\subfigure[Evolution direction of $\theta_{n,t}$]{\label{fig:fig51c}
\includegraphics[width=0.45\linewidth]{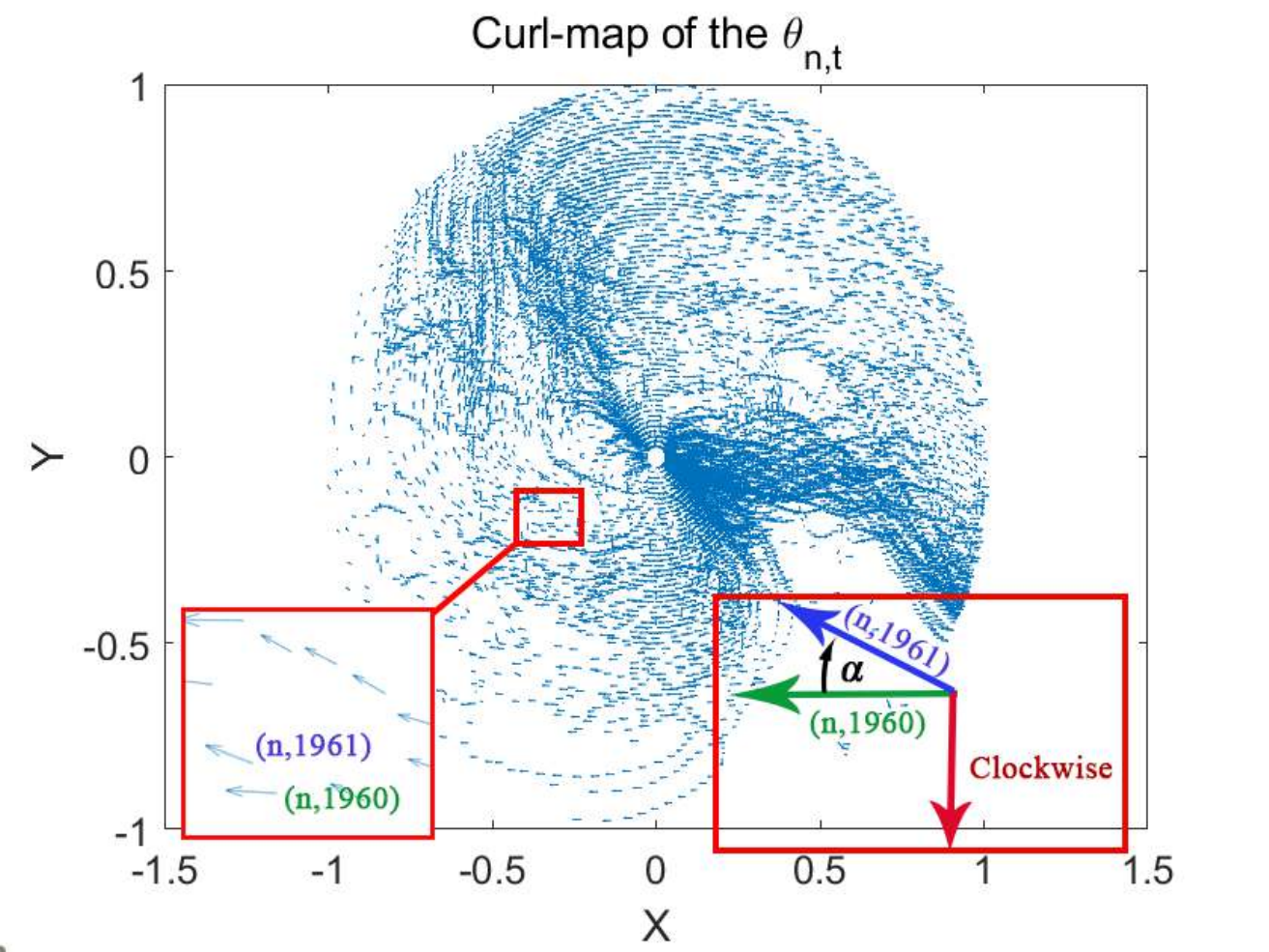}}
\subfigure[Clockwise evolution proportions]{\label{fig:fig51d}
\includegraphics[width=0.45\linewidth]{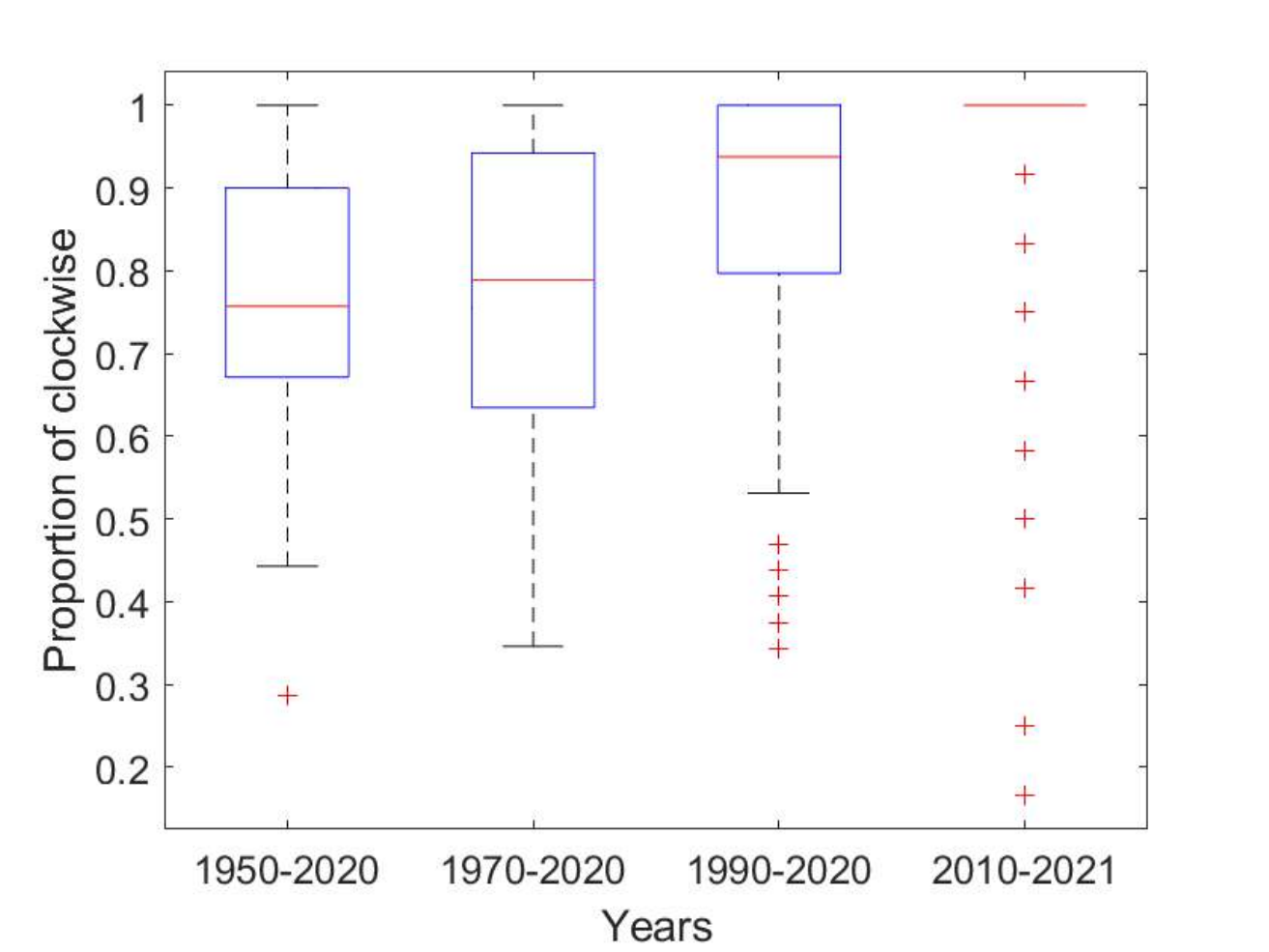}}
\caption{The alternating law of the two phenomena.}
\label{fig:fig51}
\end{figure*}

It shows the evolution of $B_{n,t}$ and $C_{n,t}$, and the indicators of each country during the period from 1950 to 2021 are represented by a series of dots from green to yellow. Taking the origin $(0,0)$ as the center, we found that the time series dots of many countries rotated clockwise around the center, that is, starting from the fourth or third quadrant, passing through the second and first phenomenon, and finally returning back to the fourth quadrant. We analyze the meaning according to the characteristics of the quadrant coefficient. The third quadrant has abundant newborns and adolescents. In this quadrant, the coefficient of the labor force and the coefficient of the elderly population are both negative. The second quadrant has prominent labor force, where the labor force coefficient is positive and the elderly coefficient is negative. The fourth quadrant has the prominent elderly population, with the positive elderly coefficient and negative labor force coefficient. The first quadrant is the transitional period from prominent labor force to prominent elderly population. The most representative countries in each quadrant and their age structures are in Appendix 3. Most evolution paths coincided with this clockwise route, with different countries having various starting and ending regions (Figure \ref{fig:fig51a}). 

This rule is more significant in polar coordinates. Figure \ref{fig:fig51b} focuses on the angle in polar coordinates as $\theta_{n,t}$, and the radius $r_{n,t}$ was uniformly quantified as an equidistant growth over time. Then, the evolution path of each country is uniformly close to the outermost circle $r=1$ over time. Here, the characteristics of clockwise evolution are more obvious. We used the vector field to analyze the evolution paths. The coordinate positions of two consecutive years form a vector, and we calculated the cross product the vector sequence. For example, the vectors of country $n$ in 1960 and 1961 were selected, and the angle $\alpha$ from $(n,1960)$ to $(n,1961)$ can be determined by the direction of their cross product. If the cross product is negative, $\alpha$ is clockwise, while if it is positive, $\alpha$ is counterclockwise (Figure \ref{fig:fig51c}).

We calculated the proportion of clockwise rotation of each country in different time periods and found that a clockwise trend was present in the evolution of most countries. Especially since the 1990s, more than 75\% of countries have experienced clockwise evolution, accounting for more than 80\% of the paths. In the last 10 years, this clockwise ratio has been very close to 100\% (Figure \ref{fig:fig51d}).

\subsection{The evolutionary trend of world population age structure}

\subsubsection{Elderly high-income and ``young'' low-income countries.}

\begin{figure}[h]
    \centering    
    \includegraphics[width=0.8\linewidth]{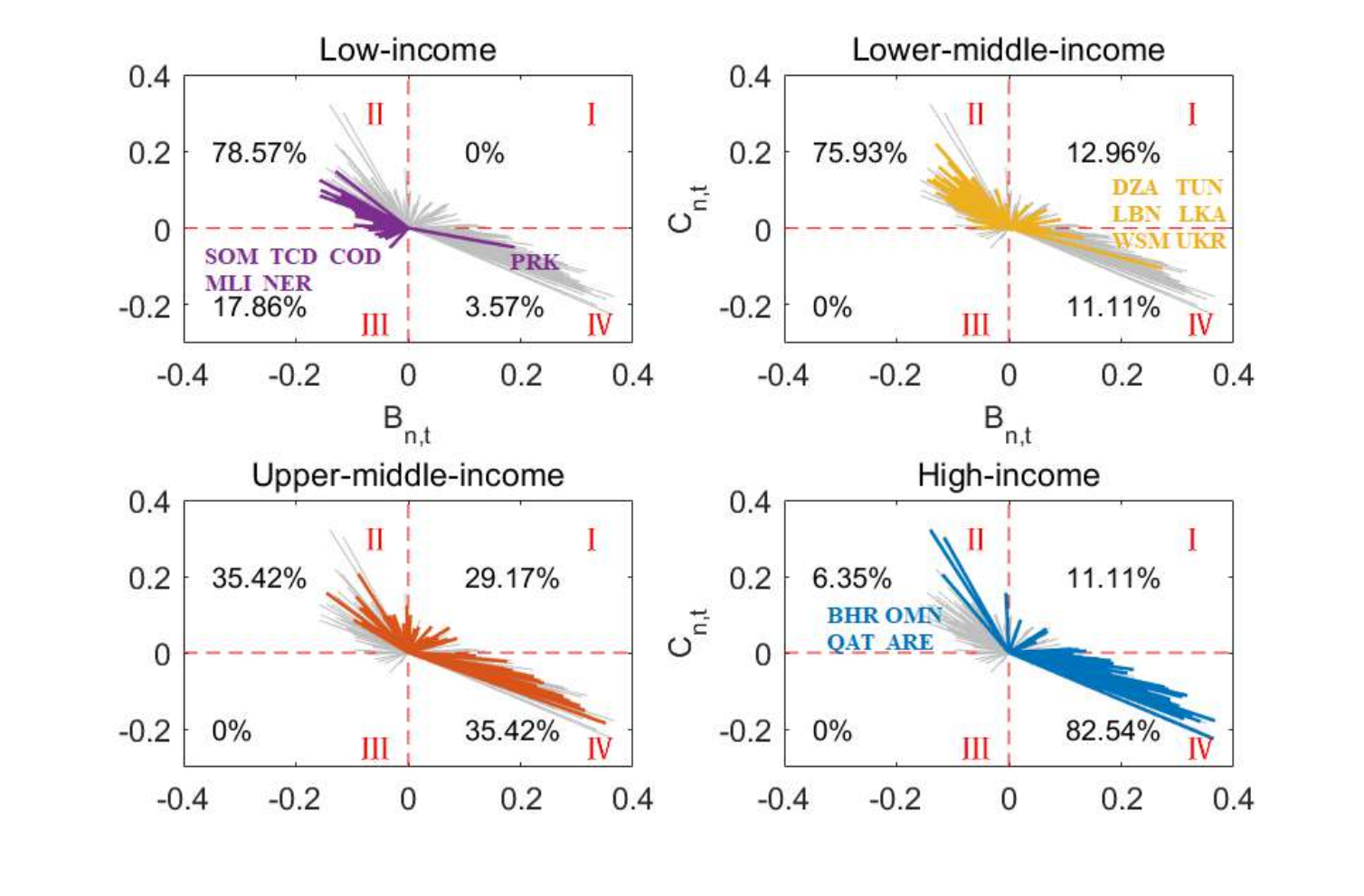}
    \caption{Characteristics of countries with different income levels in 2021-2022.}
    \label{fig:pic77}
\end{figure}

The World Bank divides countries into four categories according to their income levels, as follows: high income, upper-middle income, lower-middle income and low income. Figure \ref{fig:pic77} shows $B_{n,t}$ and $C_{n,t}$ in the period from 2021 to 2022 for each country. The gray lines connect the (0,0) and the position of countries from 2021 to 2022, and the blue, orange, yellow and purple lines represent countries in four income levels. It also shows the proportion of each category of countries located in the first to fourth quadrants.

Most low-income countries fell in the second quadrant, while a large number of high-income countries were in the fourth quadrant. The remaining two categories lie between high-income and low-income countries. For example, 75.93\% of low-middle-income countries were in the second quadrant, but 12.96\% turn clockwise to the first quadrant. For upper-middle-income countries, only 35.42\% remain in the second quadrant, with 29.17\% in the first quadrant, and another 35.42\% have been transferred to the fourth quadrant. This means that, currently, the characteristics of increasing labor structure tend to appear in relatively low-income countries, while in high-income countries, the growth trend of the elderly population is more significant.

Notably, Dem. People's Republic of Korea (PRK), as a low-income country, and Algeria (DZA), Tunisia (TUN), Lebanon (LBN), Sri Lanka (LKA), Samoa (WSM), and Ukraine (UKR), as lower-middle-income countries, are all located in the fourth quadrant and accompanied by a more prominent elderly population structure, which shows a phenomenon of ``getting old before getting rich''. The insufficient productivity and increased social burden brought about by aging will make the economic development of these countries encounter inevitable difficulties. In addition, only FIVE countries (Somalia (SOM), Chad (TCD), Democratic Republic of the Congo (COD), Mali (MLI), Niger (NER)) are in the third quadrant in 2021, which means that both the labor force and aging population in these countries are not prominent. For example, Niger (NER) is the world's youngest country, with 49.0\% of the population aged 0-14 in 2021~\citep{Niger@Misc}. The age structures of these countries and some discussions are in Appendix A.4.

There are three high-income countries still having a significantly increased proportion of the working population, including Oman (OMN), Qatar (QAT), and United Arab Emirates (ARE), which are all Gulf countries near the Persian Gulf. These countries have rich oil and natural gas resources and high income levels. The permanent foreign residents in Oman (OMN) account for approximately 50\% of the total population~\citep{Oman@Misc}, while the United Arab Emirates (ARE) and Qatar (QAT) have more than 80\% of the foreign population~\citep{ARE@Misc,Qatar@Misc}. The continuous influx of labor population makes these countries continue to have a relatively sufficient working population.

\subsubsection{Demographics on all continents are inevitably aging.}

For most countries, the higher the income level is, the more obvious the characteristics of aging, and accordingly, the characteristics of an abundant labor force are no longer significant. Since the evolution of the age structure has a certain continuity, if we do not consider the change in birth and death rates and ignore the impact of international migration, the abundance of the labor force in a certain period will naturally bring the aging trend after a period of evolution. Therefore, is it inevitable for a country to experience a path from natural population growth to a prominent labor force structure and then to an aging trend? If such a structural evolution pattern is inevitable, the world population will not continue to grow rapidly, as many scholars fear. Here, we use the prediction of the United Nations data, extend the time series of $B_{n,t}$ and $C_{n,t}$, and describe the evolution pattern of the regional age structure from 1950 to 2050.

\begin{figure}[htp]
    \centering
    \includegraphics[width=1.0\linewidth]{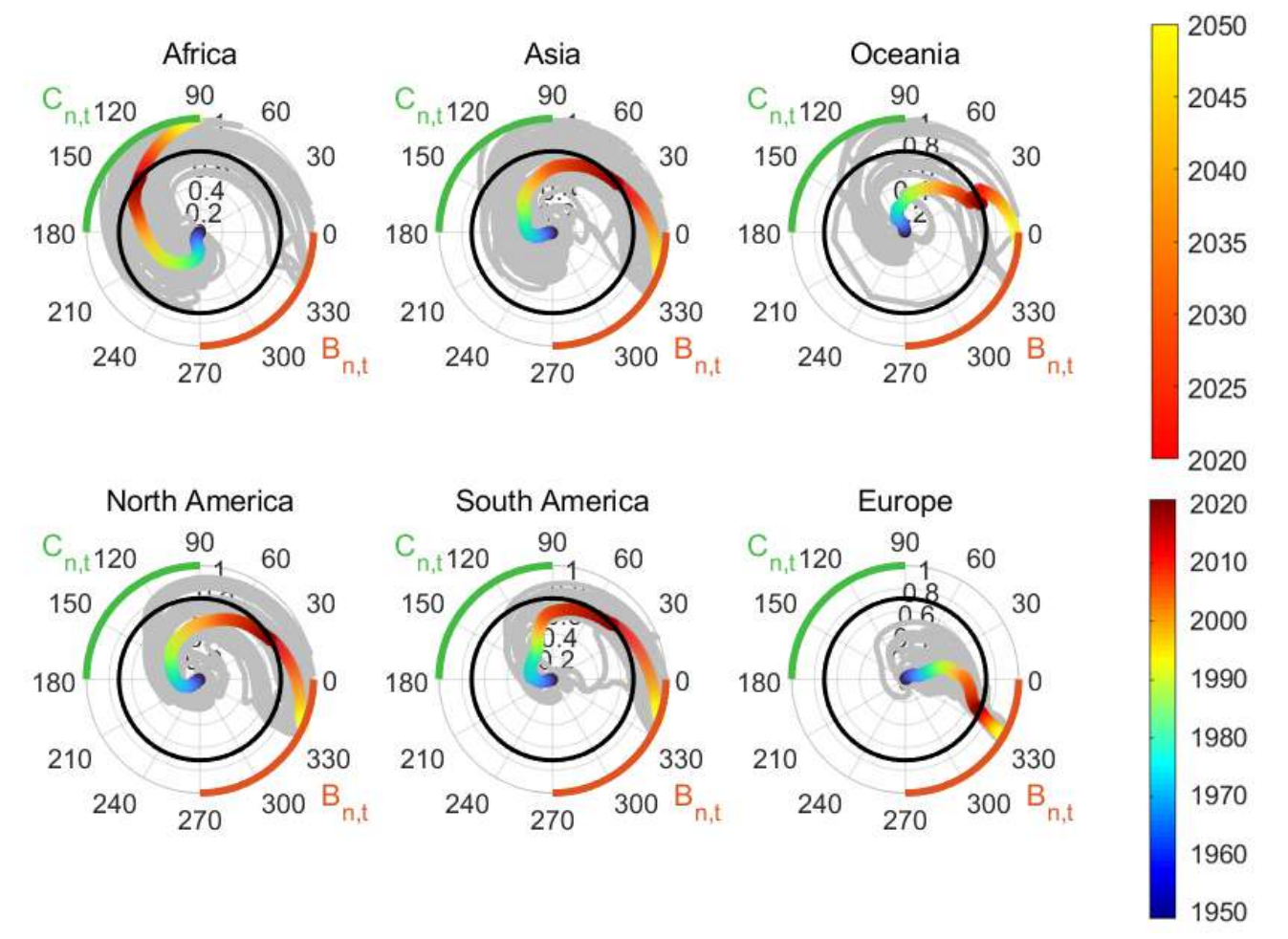}
    \caption{Evolution trend of the six continents.}
    \label{fig:pic7}
\end{figure}

Figure \ref{fig:pic7} shows the evolution trend of the age structure for the six main continents. For each continent, the gray lines are the evolution paths of all countries, and the colored lines are the paths formed by their average values, where blue to red represent empirical data and red to yellow represent forecast data from the United Nations, with the black circle serving as the dividing line. Clockwise evolution patterns generally exist in all continents within a century. For Europe, which has entered the fourth quadrant, its population structure will continue to develop in the direction of aging and labor shortages over the next 30 years. For Asia, Oceania, North America and South America, their age structure will gradually approach that of Europe, which need face the problem of aging at different times. For Asia~\citep{dlugosz2014risk,fang2015research,tey2016aging}, North~\citep{sheets2013aging} and South America~\citep{miranda2016population,angel2017aging}, their speed of entering the fourth quadrant is significantly higher than Oceania, indicating that they will face the situation of aging and labor shortage more quickly.

Currently, the age structure of Africa is still relatively young, and the working-age population is abundant. However, Africa has also shown a typical clockwise evolution trend; in addition, some African countries' angular velocities are significantly higher than those in other regions. Therefore, for Africa, if there are no significant changes in birth, death and migration in the coming decades, its proportion of the working population will continue to decline~\citep{garenne2002timing,wang2016role} and will disappear in 2050 ($\theta_{n,t}\approx 90^{\circ}$). At that time, Africa was likely to face the challenges brought by the aging trend, similar to Asia and South America a few years before. Although the total population of most countries has experienced continuous growth since World War II, by analyzing the universal path of age structure evolution in various countries, we believe that most countries will go through the development from continuous growth to abundant labor force and eventually aging, including the current ``young'' Africa, and Asia, South America with ``demographic dividend''. Even if some countries are not rich, they will still face the hidden danger of insufficient population growth and aging trends in the future.

\begin{figure}[htp]
    \centering
    \includegraphics[width=1\linewidth]{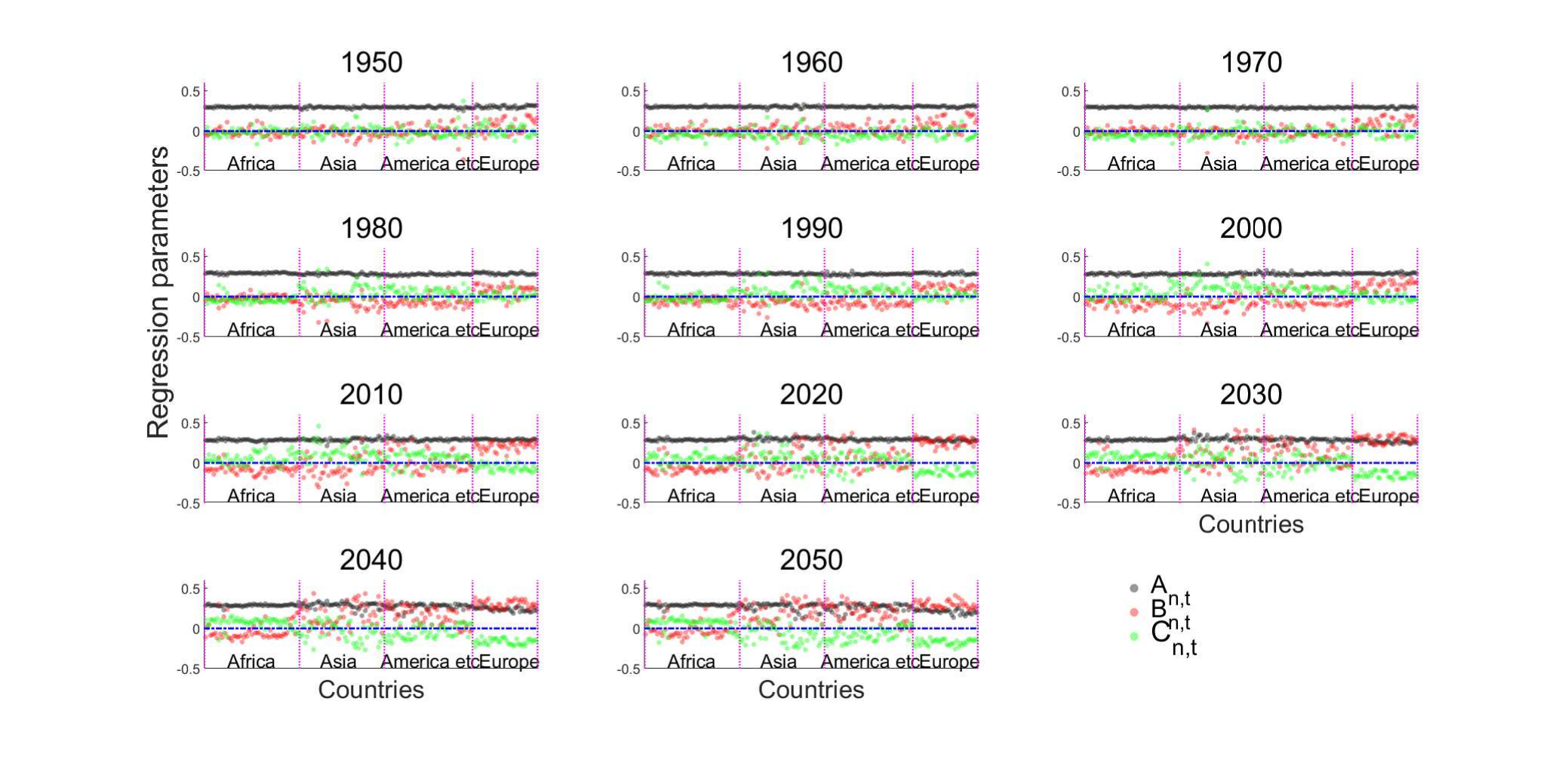}
    \caption{Evolution of three characteristics.}
    \label{fig:picS}
\end{figure}

The time series of different countries can also indicate this one-way evolution trend (Figure \ref{fig:picS}). At an interval of 10 years, we showed the evolution trend of the three age structure characteristics in various countries, as black for $A_{n,t}$, red for $B_{n,t}$ and green for $C_{n,t}$. After World War II, the characteristics of continuous population growth in most countries have not changed, which is reflected in the stability and consistency of $A_{n,t}$. However, according to the forecast data, in the future, the characteristics of sustained growth in most European, some Asian, South American and Oceanian countries will weaken, reflecting a significant decrease in $A_{n,t}$.

The characteristics of elderly growth were reflected as early as the 1950s to 1970s. Later, with the increase in the working population, some American and Asian countries showed a negative $B_{n,t}$, which means that the proportion of the elderly population has declined. From 1990 to 2020, a high proportion of the working population generally appeared in most regions except Europe, with a large number of positive $B_{n,t}$. In addition to Europe, which has already shown the characteristics of aging, most countries in Asia, South America, Oceania and even some African countries will step on the aging path after 2020. At the same time, most countries will face an insufficient labor force, with a large number of negative $C_{n,t}$.

\normalsize

\section{Conclusion}

Declining fertility, increasing longevity, and the progression of large cohorts to older ages are causing elder shares to rise throughout the world. People always tend to focus on areas where problems have occurred; for example, scholars have been concerned about aging in Europe for decades, and in recent years, they have expanded the focus to Asia and Latin America, analyzing the influence of population age structure on economy, consumption, medicine, environment, culture, and even politics. However, ``people's joy and sorrows are not connected'' (Lu Xun), and while many governments are wrestling to encourage fertility and delay aging, most African countries are still troubled by population explosion and the unemployment of young people, as in Asia several years ago. The United States census said, ``Africa is exceptionally young in 2015, and will remain so in the foreseeable future''~\citep{he2016aging}. If we are no longer limited to the traditional method of data statistics but interpret the evolution of the age structure in the past 72 years, such an optimistic judgment is worth pondering.

In this paper, we suggest a complex system described by age-specific data from 200 countries during 1950-2021, which constitutes a complex tensor with three dimensions: time, country and age group. Eigen microstate theory, originating from statistical physics, does not reduce the dimension of high-dimensional tensors as a traditional statistical method but explores some macro phenomena and evolution laws of the system and the collective behaviors of objects by integrating information from different dimensions. Here, age structure is the microstate of each country. The changes in these individual microstates over time could emerge from the macrostate evolution of the entire global population system.

First, it finds three main characteristics of global population evolution in the past 72 years and restores most information in the original age-specific data with three sets of eigenvectors. That is, 88.66\% of the information described by 302,400 values is reproduced with just 879 values. It shows that after World War II, the population of most countries continued to grow, but the growth rates had different slowdowns (as EM1). Since the 1950s, the world has been evolving toward aging.

In addition, the analysis including the microstate of each country will show more evolution laws in addition to the current aging situation. The iteration of the population age structure has inertia, which presents the growth and coexistence of the working population and the elderly population. Here, it constructs a space composed of two macrostates, the prominent working population (EM3) and the prominent aging population (EM2). Then, we could compare the microcosmic state of countries by their position in space at different times and describe the macroscopic evolution law of the whole system. Recently, 82.14\% of low-income countries were located in the EM3 area, and 85.71\% of high-income countries were located in the EM2 area. For the remaining two categories, 64.81\% of low-middle-income countries and 18.75\% of upper-middle-income countries were in the EM3 area, and 9.26\% of low-middle-income countries and 35.42\% of upper-middle-income countries were located in the EM2 area. It shows that with economic growth, the country's population structure has a universal rule of transition from a sufficient labor force to an aging population.

Finally, it draws the evolutionary path of countries in the two macrostate spaces, where most paths turn clockwise, and the prominence of the age structure ranges from the newborn population to the working population and then to the aging population. The World Bank's forecast data extended these evolution paths to 2050. The inevitable trend of aging will replace the slow growth of the population and become the first macrostate of world population evolution, which indicates that the demographics on all continents are inevitably aging, at present or in the near future, including the current "young" Africa, and Asia, South America with a "demographic dividend".

\section*{Acknowledgments}
\renewcommand{\thefigure}{A\arabic{figure}}
\renewcommand{\thetable}{A\arabic{table}}
\setcounter{figure}{0}
\setcounter{table}{0}

The data used in this study are provided by the Big Data Center of the School of System Science, Beijing Normal University (\url{https://sssdata.bnu.edu.cn/}). We appreciate the comments and helpful suggestions from Professors Dahui Wang, Honggang Li, Handong Li and Zengru Di, Dr. Yu Shun and Mr. Teng Liu. This work was supported by the Chinese National Social Science Foundation (22BRK021), Humanities and Social Sciences Foundation of the Ministry of Education of China (20YJAZH010) and Interdisciplinary Construction Project of Beijing Normal University.

\clearpage


\providecommand{\newblock}{}

\clearpage
\appendix
\section*{Appendix}

\renewcommand{\theequation}{A.\arabic{equation}}
\renewcommand{\thefigure}{A.\arabic{figure}}
\renewcommand{\thesubsection}{A.\arabic{subsection}}

\subsection{Validity of decomposed results}

\subsubsection{Determination of rank.}
For each rank we randomize the initial value 500 times and choose the final decomposition result according to the fit value. Since the fit values obtained by some random initial values are very close, when we select the optimal result of each rank, in addition to considering the size of the fit value, we also consider the orthogonality index, and hope that each EM explains the information is as varied as possible. The orthogonality index of EM$_i$ and EM$_j$ can be calculated by inner product, its form is $Z(i,j) = \left| { < \bm{a_i},\bm{a_j} > } \right| * \left| { < \bm{c_i},\bm{c_j} > } \right| * \left| { < \bm{t_i},\bm{t_j} > } \right|$.
The expression to measure the overall orthogonality of the decomposition results is as follows, 

\begin{equation}
	Z = \sum\limits_{i = 1}^R \sum\limits_{j = i + 1}^R {\left| { < \bm{a_i},\bm{a_j} > } \right| * \left| { < \bm{c_i},\bm{c_j} > } \right| * \left| { < \bm{t_i},\bm{t_j} > } \right|}.
	\label{eq:fl_rank}
\end{equation}

The lower the orthogonality index, the higher the orthogonality between EMs in the decomposition results, and the more different interpretation angles of each EM will be. We set the random initial value 500 times under the given rank,  and consider the decomposition result with the fit value in the top 10\% and the lowest orthogonality index as the optimal one. It is worth noting that this choice will not lose the fit value. The difference between the highest fit value and the fit value considered to be orthogonal does not exceed 1\%, and the difference under some ranks is 0.

\begin{figure}[htp]
    \centering
    \includegraphics[width=0.6\linewidth]{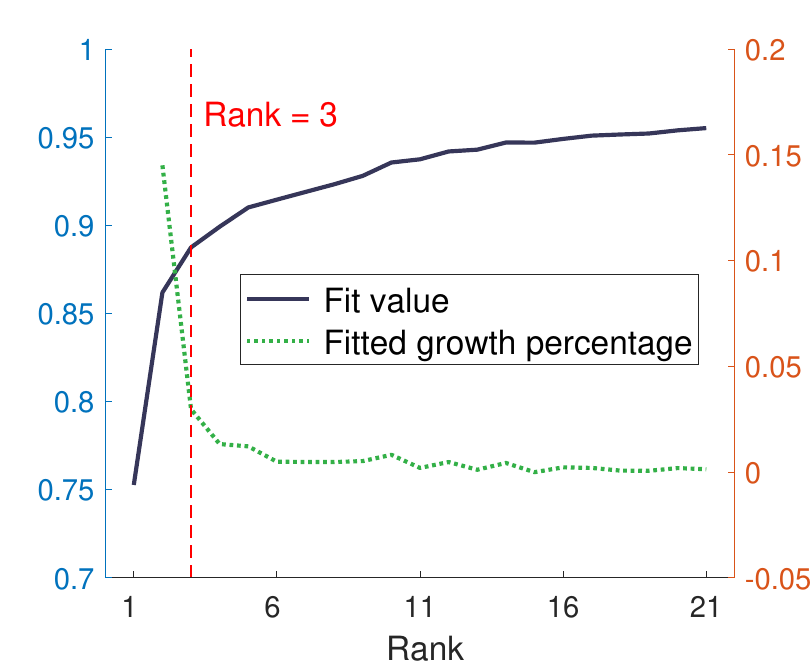}
    \caption{Rank selection based on fit value.}
    \label{fig:pic881}
\end{figure}

As shown in the Figure \ref{fig:pic881}, we drawed the fit value $f$ under different ranks. The two curves are the fit value $f$ and the percentage of growth $\frac{{f(n) - f(n - 1)}}{{f(n - 1)}}$ (where $n$ is the rank value, $n\geq 2$). It shows that when the rank value is greater than 3, a significant increase in the rank does not lead to a significant increase in the fit value.



\begin{figure}[htp]
    \centering
    \includegraphics[width=1\linewidth]{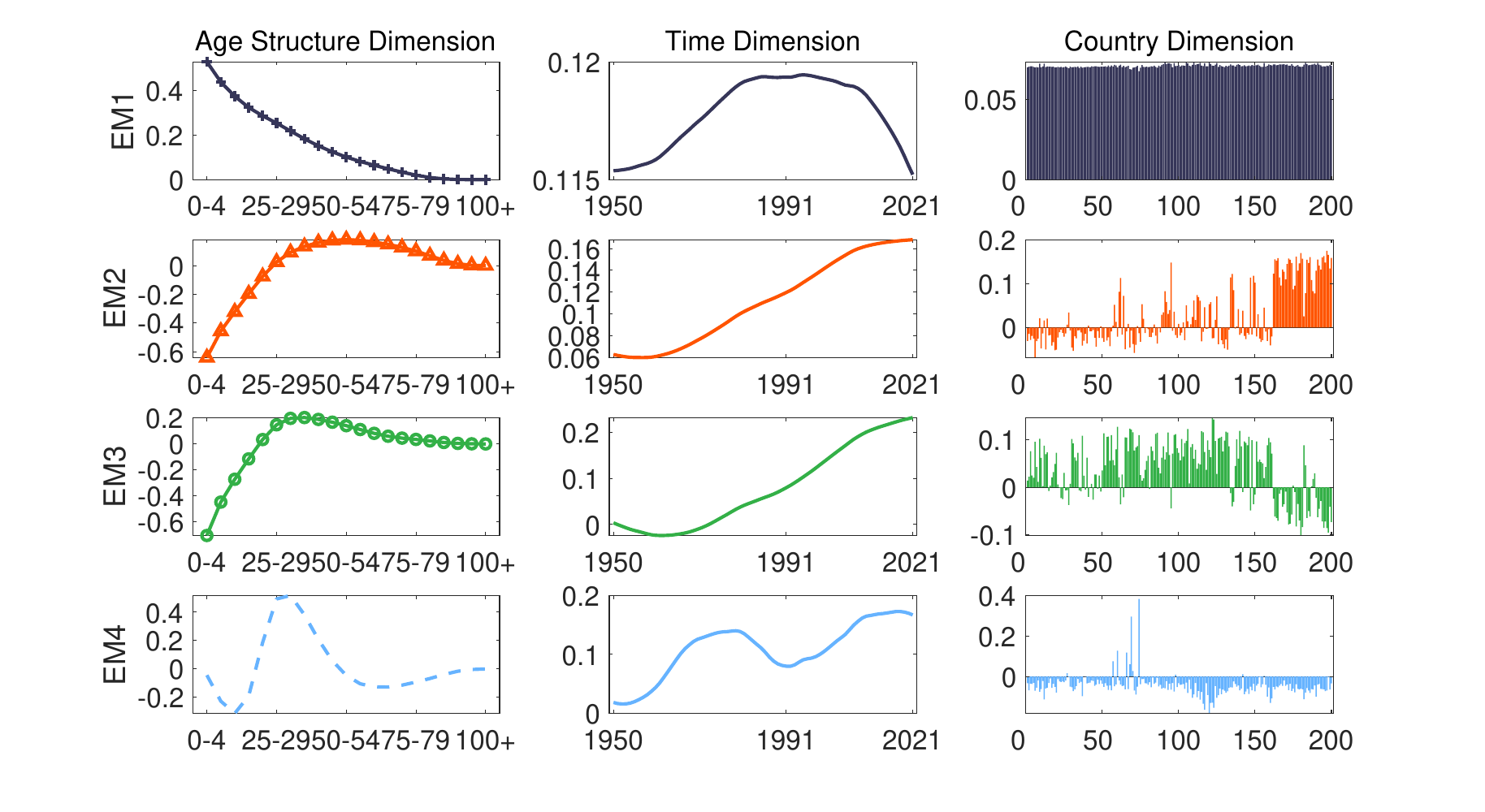}
    \caption{Decomposition result}
    \label{fig:pic81}
\end{figure}

Furthermore, we plot the decomposition results with rank=4 (Figure \ref{fig:pic81}). The result is consistent with the that of rank=3 in the first three EMs, highlighting the overall trend, aging, and labor force respectively, and the values of most countries are more prominent in the country dimension. However, the value of the national dimension of EM4 is reflected in individual countries, and it is not a general rule that most countries have. In summary, we choose the rank=3 here.

\subsubsection{The results of 10 experiments.}

\begin{figure}[htp]
    \centering
    \includegraphics[width=1\linewidth]{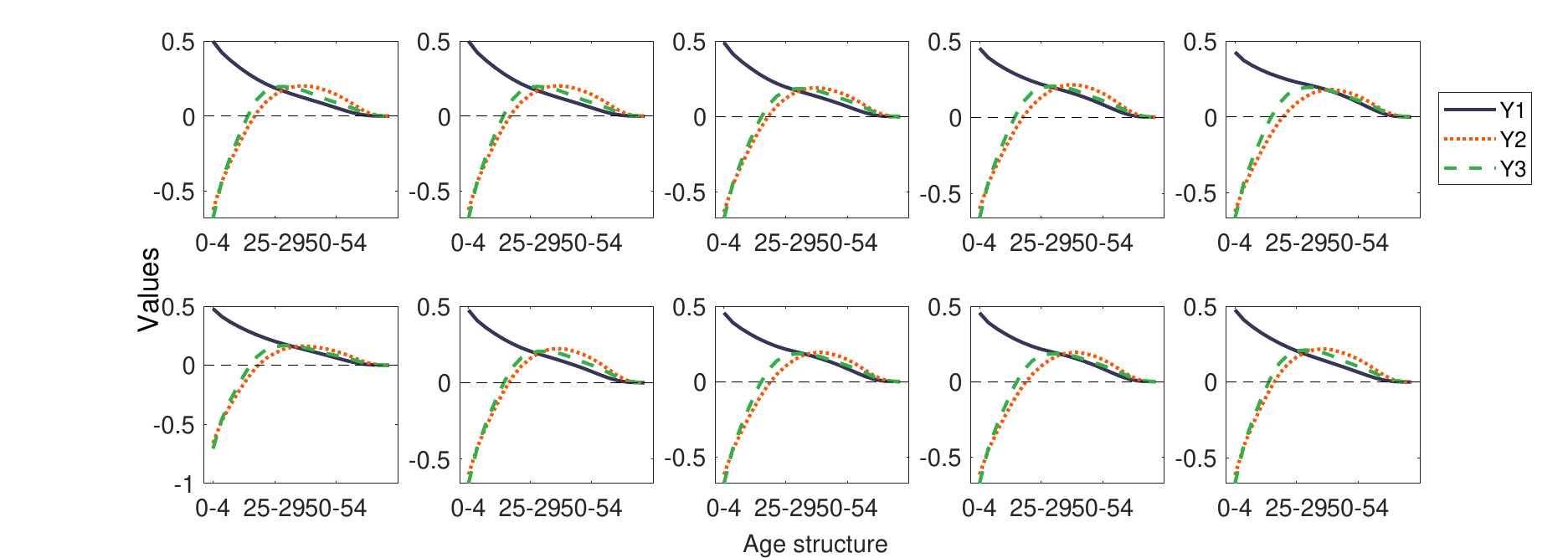}
    \caption{Select the result of the maximum fit value}
    \label{fig:pic_fl1}
\end{figure}

Based on rank=3, we conducted 10 experiments, and each experiment randomly assigned 500 initial values. The decomposition results corresponding to the best fitting value obtained in each experiment are as follows. Figure \ref{fig:pic_fl1} shows the age structure of the decomposition results. 

\begin{figure}[htp]
    \centering
    \includegraphics[width=1\linewidth]{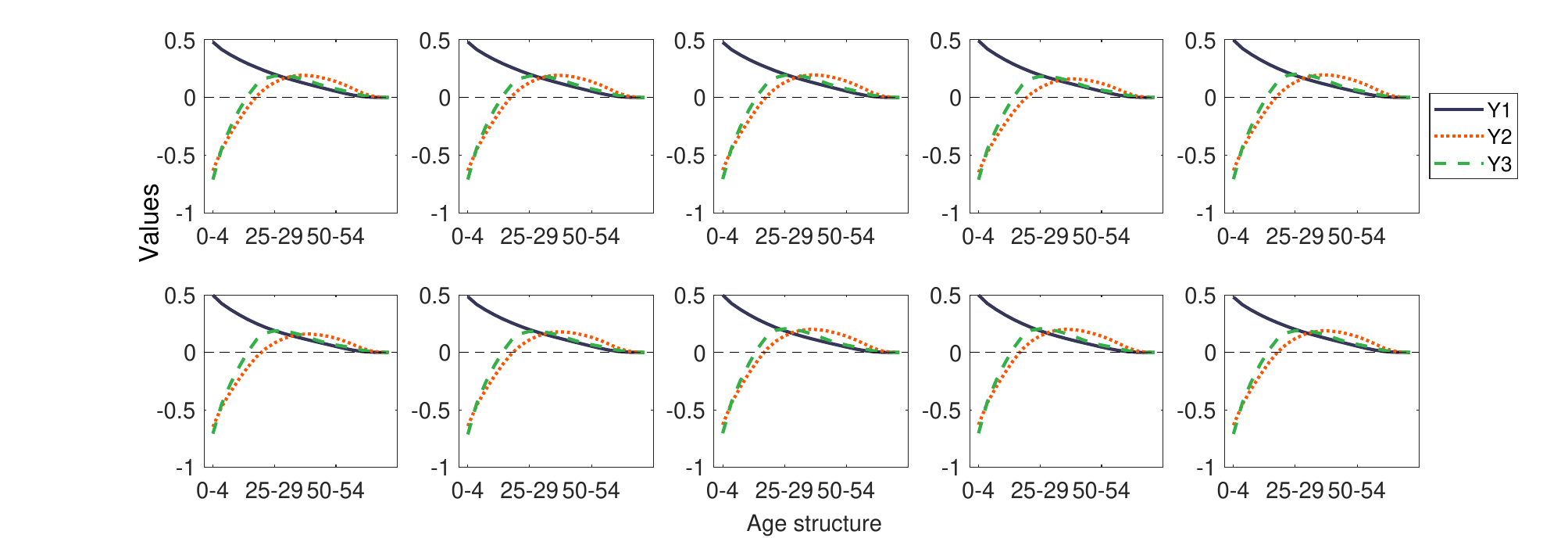}
    \caption{Select the top 10\% fit value and orthogonal results}
    \label{fig:picfl2}
\end{figure}

We also considered the orthogonality index such that the three age structures obtained from the decomposition results should be as orthogonal as possible. The following is the result of each experiment with fit value ranking in the top 10\% and orthogonal index minimum, Figure \ref{fig:picfl2} shows the age structure of the decomposition results. In both cases, the revealed characteristics of age structure were similar. The two major evolutionary trends are as follows: prominent aging and prominent growth in the labor force. We obtained similar and stable results in these experiments.

In summary, we sacrificed less than 1\% (fit value comparison) of the fit value for a result with a low orthogonality index, see Table \ref{tab:tabfl2} for details.
\begin{table}[!ht]
\tiny
    \centering
    \caption{Fit value and orthogonality index}
    \label{tab:tabfl2}
    \begin{tabular}{|l|l|l|l|l|l|l|l|l|l|l|}
    \hline
        Experiment number & 1 & 2 & 3 & 4 & 5 & 6 & 7 & 8 & 9 & 10 \\
        \hline
         maximum fit value & 0.8873  & 0.8873  & 0.8875  & 0.8873  & 0.8874  & 0.8874  & 0.8875  & 0.8874  & 0.8874  & 0.8874  \\ \hline
         Corresponding Orthogonality Index & 1.0549  & 1.0549  & 1.4753  & 1.2451  & 1.3315  & 1.1832  & 1.3087  & 1.4102  & 1.4102  & 1.1008 \\ 
        \hline
        Orthogonal - fit values & 0.8865  & 0.8865  & 0.8864  & 0.8865  & 0.8867  & 0.8865  & 0.8866  & 0.8868  & 0.8868  & 0.8865  \\ \hline
        Corresponding Orthogonality Index  & 0.6007  & 0.6007  & 0.5812  & 0.6203  & 0.6096  & 0.5972  & 0.6059  & 0.6429  & 0.6429  & 0.5780 \\ \hline
    \end{tabular}
    
\end{table}

\subsubsection{Other decomposition methods.}

\begin{figure}[h]
    \centering
    \subfigure[HOSVD]{
    \label{fig:figure1-a} 
    \includegraphics[width=0.45\textwidth]{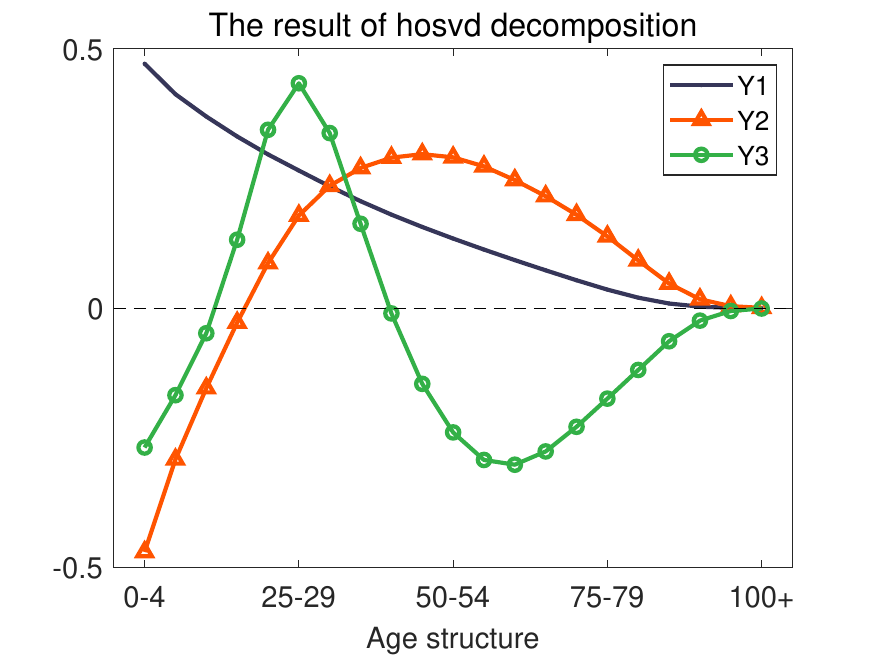}}
    \subfigure[Characteristics of population age structure evolution]{
    \label{fig:figure1-b} 
    \includegraphics[width=0.45\textwidth]{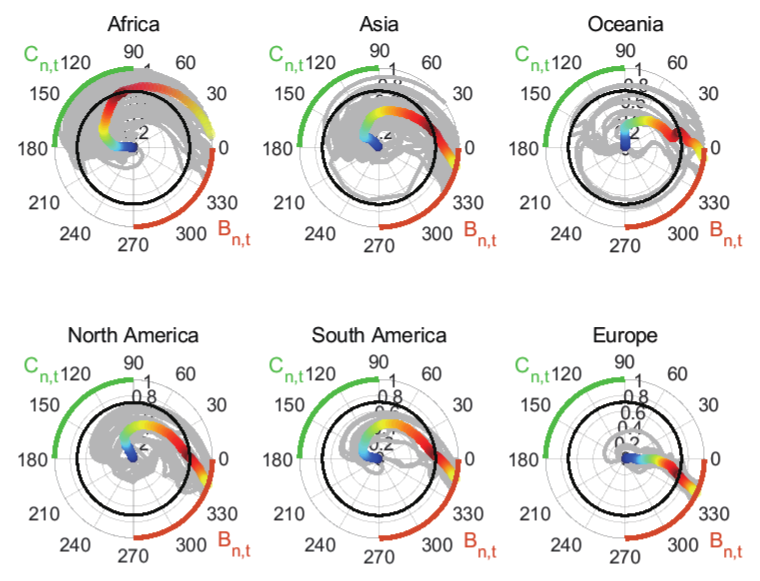}}
    \caption{HOSVD decomposition methods}
    \label{fig:pic34}
\end{figure}

We also used HOSVD and other decomposition methods and found that the age structure of EM1 is still the overall trend, and the age structures of EM2 and EM3 had maximum values in the elderly and labor ages, respectively. The difference is that the HOSVD decomposition results showed that the maximum age proportion of EM3 was more prominent than that obtained by CP decomposition. However, this will not affect the universal evolution law of the age structure we proposed (as in Figure \ref{fig:pic34}).

\subsection{Country dimensions for decomposition results.}

\begin{figure}[h]
    \centering
    \subfigure[$\bm{c_2}$]{
    \label{fig:figure2-a} 
    \includegraphics[width=1\textwidth]{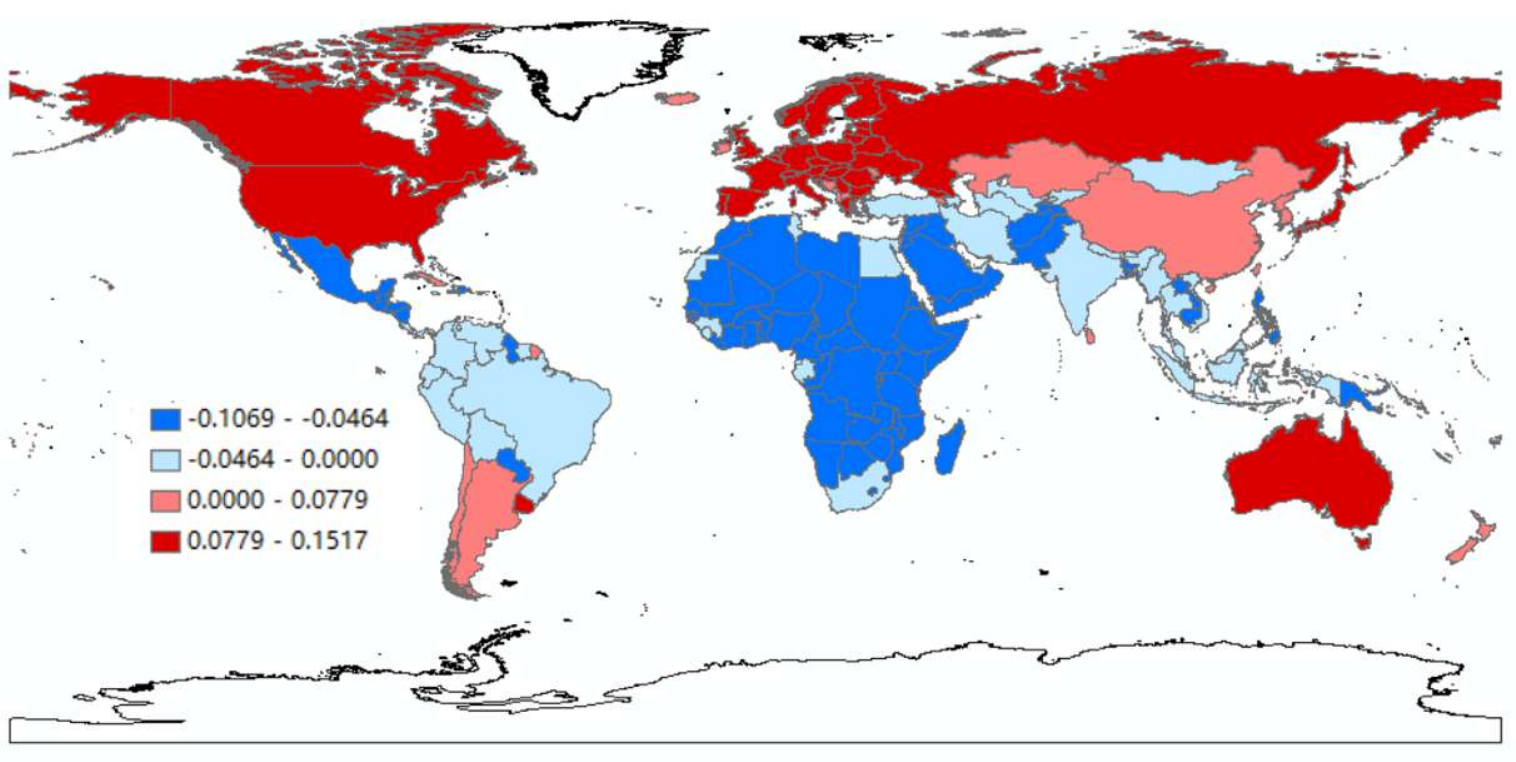}}
    \subfigure[$\bm{c_3}$]{
    \label{fig:figure2-b} 
    \includegraphics[width=1\textwidth]{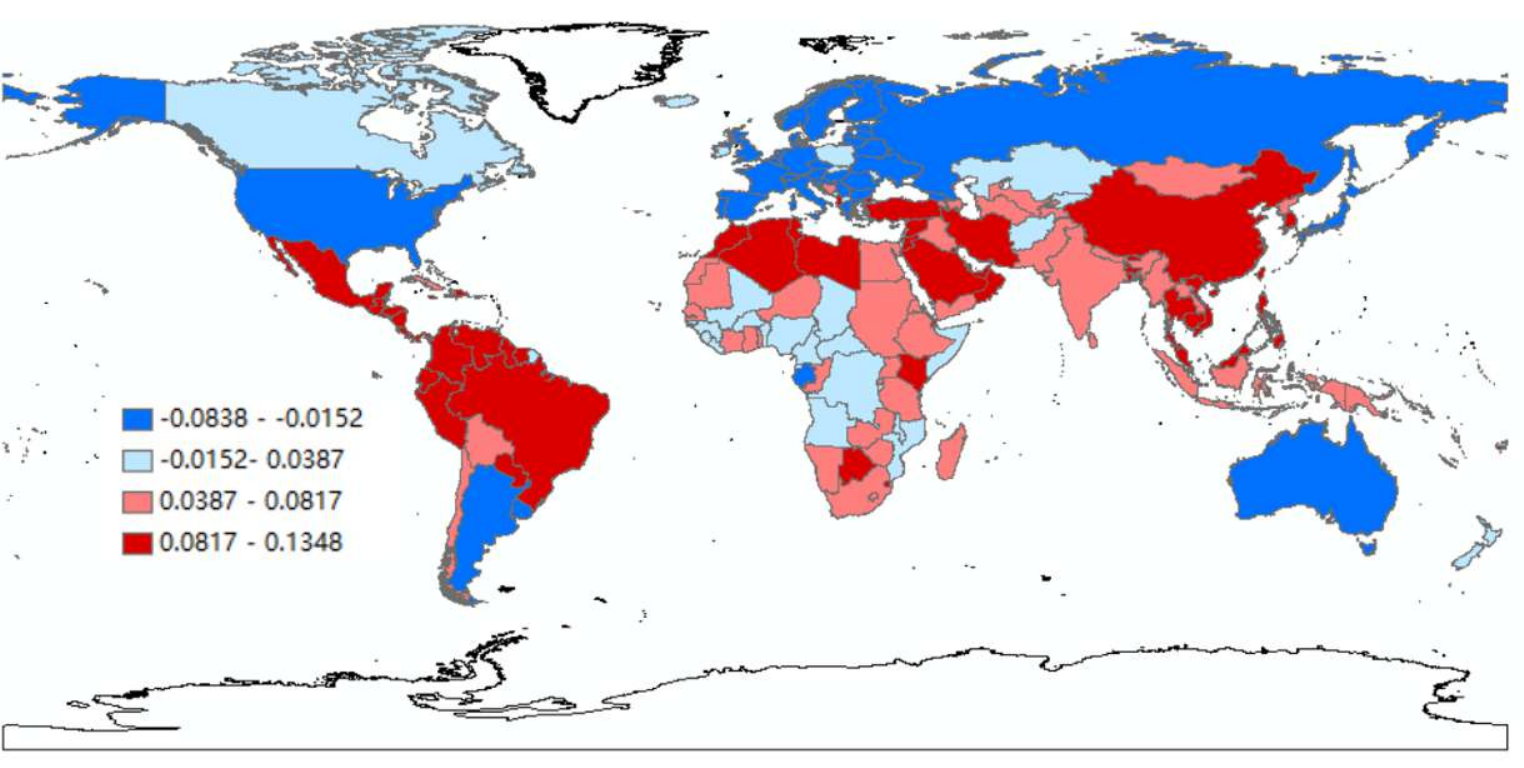}}
    \caption{Country dimension of EM2 and EM3}
    \label{fig:pic35}
\end{figure}

In Figure \ref{fig:pic35}, the values of $\bm{c_2}$ and $\bm{c_3}$ in each country are shown separately. The countries corresponding to the positive value of $\bm{c_2}$ are mainly concentrated in Europe, and the countries corresponding to the positive value of $\bm{c_3}$ are mainly concentrated in Africa, Asia and other regions. 

\subsection{The coefficient of the aging trend will keep on increasing in the next 20 years.}

\begin{figure}[htp]
    \centering
    \includegraphics[width=0.6\linewidth]{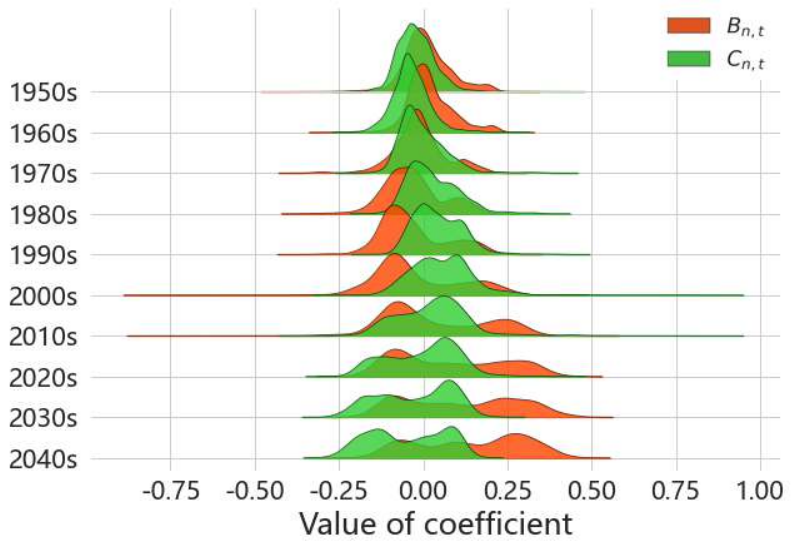}
    \caption{Ridge map with forecast datas}
    \label{fig:pic88}
\end{figure}

Figure \ref{fig:pic88} is a supplement to Figure \ref{fig:fig55b} with forecast data from the 2020s to the 2040s. Here, the ridgeplot shows the distribution and evolution of $B_{n,t}$ and $C_{n,t}$. The distributions of $B_{n,t}$ and $C_{n,t}$ gradually evolved from unimodal to bimodal distributions, reflecting the trend of countries' demographic characteristics from similarly to polarized. For $B_{n,t}$ (red), the peak on the right gradually replaced the peak on the left to become the most frequent one, meaning the aging trend will continue to increase in the next 30 years, and this phenomenon will occur in most countries.

\subsection{Age structure characteristics of representative countries in four quadrants.}

\begin{figure}[htp]
    \centering
     \subfigure[Countries in the four quadrants]{    \includegraphics[width=1\linewidth]{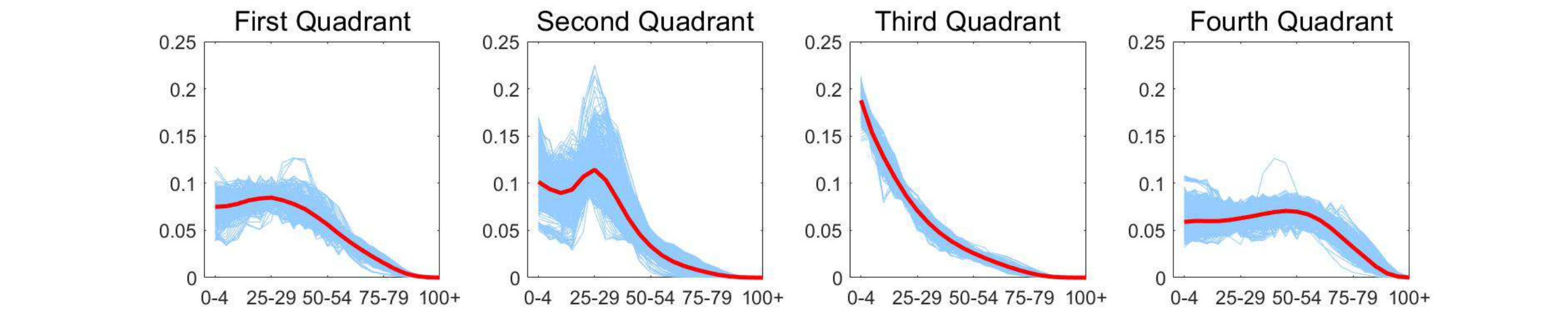}}
    \subfigure[Countries of different income-levels]{\includegraphics[width=0.8\linewidth]{pic_pdf/ch33_enconomy}}
    \caption{The age structure represented in the four quadrants}
    \label{fig:pic9}
\end{figure}

Figure \ref{fig:pic9} (a) shows the age structure characteristics of some countries, with blue lines describing the age structures for each country and red lines representing their mean value. With $B_{n,t}$ and $C_{n,t}$ as horizontal and vertical coordinates, the positions of countries fall in the first to fourth quadrants (Figure \ref{fig:pic9} (b)). The second quadrant (II) has the prominent characteristics of the working-age population structure, and most low-income (78.57\%) and lower-middle-income (75.93\%) countries belong to this type. The third quadrant (III) has the prominent characteristics of the underage population structure, including only four low-income countries, and these countries have more than 46.53\% of the population under the age of 14. The fourth quadrant (IV) has prominent characteristics of the elderly population structure, including most high-income countries (82.54\%).

\subsection{Evolution of age structure in some countries.}

\begin{figure}[htp]
    \centering
    \includegraphics[width=0.3\linewidth]{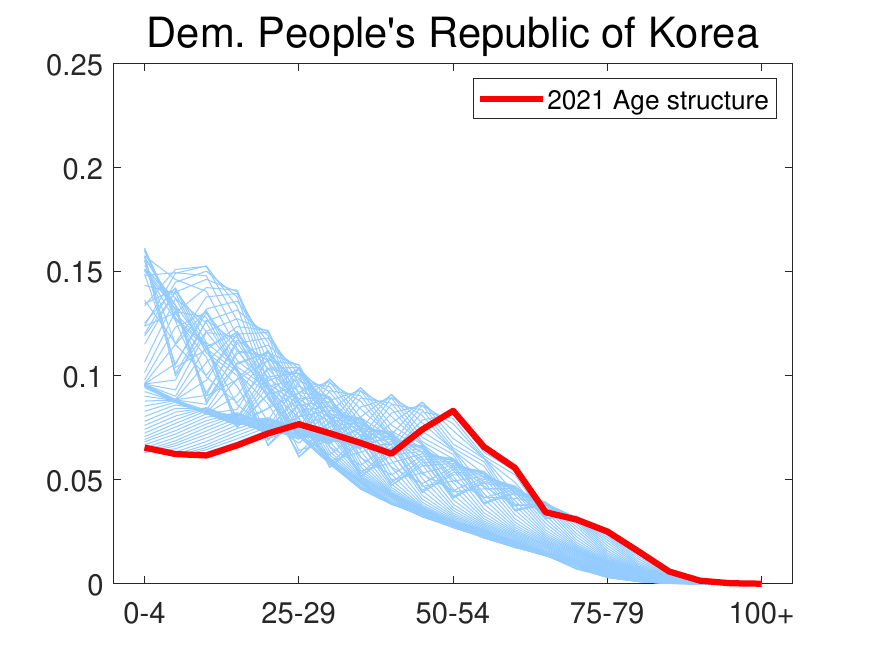}
    \caption{Dem. People's Republic of Korea Age Structure}
    \label{fig:pic10}
\end{figure}

\begin{figure}[htp]
    \centering
    \includegraphics[width=1\linewidth]{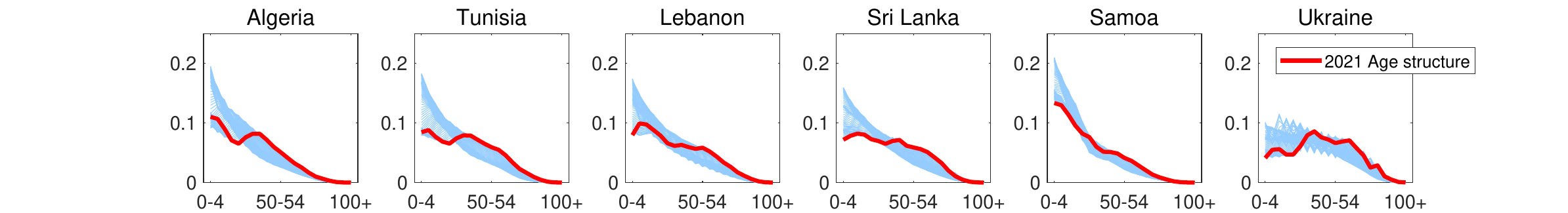}
    \caption{Some Lower-middle-income economies}
    \label{fig:pic11}
\end{figure}

Dem. People's Republic of Korea (PRK) is the only low-income country with an aging trend (Figure \ref{fig:pic10}), and Figure \ref{fig:pic11} presents some ageing countries at the low-middle-income level. Here, the blue lines represent the age structure of these countries from 1950 to 2020, and the red line represents their age structure of 2021.

\begin{figure}[htp]
    \centering
    \includegraphics[width=1\linewidth]{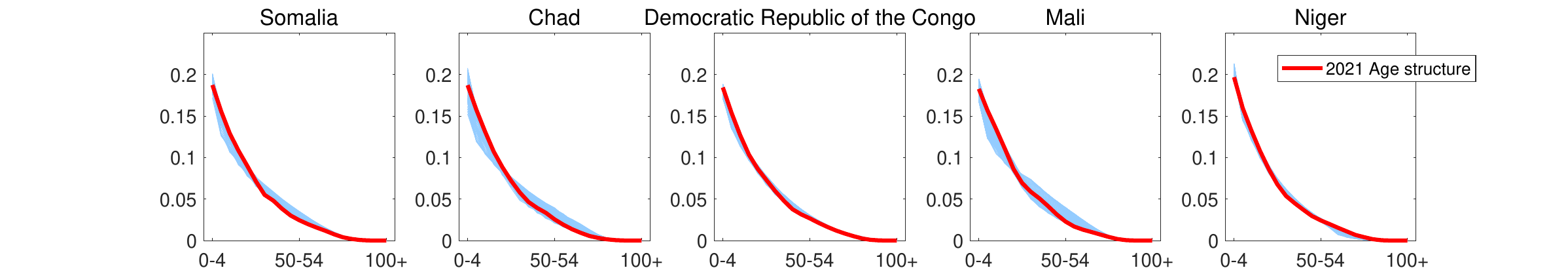}
    \caption{Countries in the third quadrant.}
    \label{fig:pic12}
\end{figure}

Figure \ref{fig:pic12} depicts the only five countries in the third quadrant in 2021 with prominent characteristics of underage population structure, including Somalia(SOM), Chad(TCD), Democratic Republic of the Congo(COD), Mali(MLI) and Niger(NER).

\begin{figure}[htp]
    \centering
    \includegraphics[width=1\linewidth]{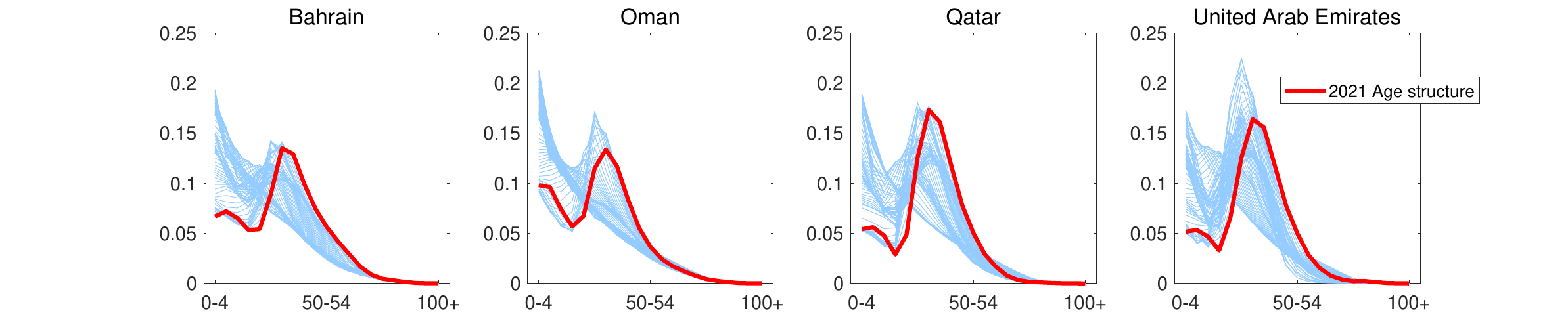}
    \caption{High-income countries in the second quadrant.}
    \label{fig:pic13}
\end{figure}

Figure \ref{fig:pic13} shows the three high-income countries in the second quadrant with prominent characteristics of the workforce, including Oman(OMN), Qatar(QAT) and the United Arab Emirates(ARE).

\subsection{List of countries}

\begin{longtable}{|l|l|l|l|}
 
    \hline
        No. & Countries or regions & ISO3 & Continent \\ 
        \endhead
        \hline
        1 & Burundi & BDI & Africa \\ \hline
        2 & Comoros & COM & Africa \\ \hline
        3 & Djibouti & DJI & Africa \\ \hline
        4 & Eritrea & ERI & Africa \\ \hline
        5 & Ethiopia & ETH & Africa \\ \hline
        6 & Kenya & KEN & Africa \\ \hline
        7 & Madagascar & MDG & Africa \\ \hline
        8 & Malawi & MWI & Africa \\ \hline
        9 & Mauritius & MUS & Africa \\ \hline
        10 & Mayotte & MYT & Africa \\ \hline
        11 & Mozambique & MOZ & Africa \\ \hline
        12 & Réunion & REU & Africa \\ \hline
        13 & Rwanda & RWA & Africa \\ \hline
        14 & Seychelles & SYC & Africa \\ \hline
        15 & Somalia & SOM & Africa \\ \hline
        16 & South Sudan & SSD & Africa \\ \hline
        17 & Uganda & UGA & Africa \\ \hline
        18 & United Republic of Tanzania & TZA & Africa \\ \hline
        19 & Zambia & ZMB & Africa \\ \hline
        20 & Zimbabwe & ZWE & Africa \\ \hline
        21 & Angola & AGO & Africa \\ \hline
        22 & Cameroon & CMR & Africa \\ \hline
        23 & Central African Republic & CAF & Africa \\ \hline
        24 & Chad & TCD & Africa \\ \hline
        25 & Congo & COG & Africa \\ \hline
        26 & Democratic Republic of the Congo & COD & Africa \\ \hline
        27 & Equatorial Guinea & GNQ & Africa \\ \hline
        28 & Gabon & GAB & Africa \\ \hline
        29 & Sao Tome and Principe & STP & Africa \\ \hline
        30 & Botswana & BWA & Africa \\ \hline
        31 & Eswatini & SWZ & Africa \\ \hline
        32 & Lesotho & LSO & Africa \\ \hline
        33 & Namibia & NAM & Africa \\ \hline
        34 & South Africa & ZAF & Africa \\ \hline
        35 & Benin & BEN & Africa \\ \hline
        36 & Burkina Faso & BFA & Africa \\ \hline
        37 & Cabo Verde & CPV & Africa \\ \hline
        38 & Côte d'Ivoire & CIV & Africa \\ \hline
        39 & Gambia & GMB & Africa \\ \hline
        40 & Ghana & GHA & Africa \\ \hline
        41 & Guinea & GIN & Africa \\ \hline
        42 & Guinea-Bissau & GNB & Africa \\ \hline
        43 & Liberia & LBR & Africa \\ \hline
        44 & Mali & MLI & Africa \\ \hline
        45 & Mauritania & MRT & Africa \\ \hline
        46 & Niger & NER & Africa \\ \hline
        47 & Nigeria & NGA & Africa \\ \hline
        48 & Senegal & SEN & Africa \\ \hline
        49 & Sierra Leone & SLE & Africa \\ \hline
        50 & Togo & TGO & Africa \\ \hline
        51 & Algeria & DZA & Africa \\ \hline
        52 & Egypt & EGY & Africa \\ \hline
        53 & Libya & LBY & Africa \\ \hline
        54 & Morocco & MAR & Africa \\ \hline
        55 & Sudan & SDN & Africa \\ \hline
        56 & Tunisia & TUN & Africa \\ \hline
        57 & Western Sahara & ESH & Africa \\ \hline
        58 & Armenia & ARM & Asia \\ \hline
        59 & Azerbaijan & AZE & Asia \\ \hline
        60 & Bahrain & BHR & Asia \\ \hline
        61 & Cyprus & CYP & Asia \\ \hline
        62 & Georgia & GEO & Asia \\ \hline
        63 & Iraq & IRQ & Asia \\ \hline
        64 & Israel & ISR & Asia \\ \hline
        65 & Jordan & JOR & Asia \\ \hline
        66 & Kuwait & KWT & Asia \\ \hline
        67 & Lebanon & LBN & Asia \\ \hline
        68 & Oman & OMN & Asia \\ \hline
        69 & Qatar & QAT & Asia \\ \hline
        70 & Saudi Arabia & SAU & Asia \\ \hline
        71 & State of Palestine & PSE & Asia \\ \hline
        72 & Syrian Arab Republic & SYR & Asia \\ \hline
        73 & Turkey & TUR & Asia \\ \hline
        74 & United Arab Emirates & ARE & Asia \\ \hline
        75 & Yemen & YEM & Asia \\ \hline
        76 & Kazakhstan & KAZ & Asia \\ \hline
        77 & Kyrgyzstan & KGZ & Asia \\ \hline
        78 & Tajikistan & TJK & Asia \\ \hline
        79 & Turkmenistan & TKM & Asia \\ \hline
        80 & Uzbekistan & UZB & Asia \\ \hline
        81 & Afghanistan & AFG & Asia \\ \hline
        82 & Bangladesh & BGD & Asia \\ \hline
        83 & Bhutan & BTN & Asia \\ \hline
        84 & India & IND & Asia \\ \hline
        85 & Iran (Islamic Republic of) & IRN & Asia \\ \hline
        86 & Maldives & MDV & Asia \\ \hline
        87 & Nepal & NPL & Asia \\ \hline
        88 & Pakistan & PAK & Asia \\ \hline
        89 & Sri Lanka & LKA & Asia \\ \hline
        90 & China & CHN & Asia \\ \hline
        91 & China, Hong Kong SAR & HKG & Asia \\ \hline
        92 & China, Macao SAR & MAC & Asia \\ \hline
        93 & China, Taiwan Province of China & TWN & Asia \\ \hline
        94 & Dem. People's Republic of Korea & PRK & Asia \\ \hline
        95 & Japan & JPN & Asia \\ \hline
        96 & Mongolia & MNG & Asia \\ \hline
        97 & Republic of Korea & KOR & Asia \\ \hline
        98 & Brunei Darussalam & BRN & Asia \\ \hline
        99 & Cambodia & KHM & Asia \\ \hline
        100 & Indonesia & IDN & Asia \\ \hline
        101 & Lao People's Democratic Republic & LAO & Asia \\ \hline
        102 & Malaysia & MYS & Asia \\ \hline
        103 & Myanmar & MMR & Asia \\ \hline
        104 & Philippines & PHL & Asia \\ \hline
        105 & Singapore & SGP & Asia \\ \hline
        106 & Thailand & THA & Asia \\ \hline
        107 & Timor-Leste & TLS & Asia \\ \hline
        108 & Viet Nam & VNM & Asia \\ \hline
        109 & Antigua and Barbuda & ATG & North America \\ \hline
        110 & Aruba & ABW & North America \\ \hline
        111 & Bahamas & BHS & North America \\ \hline
        112 & Barbados & BRB & North America \\ \hline
        113 & Cuba & CUB & North America \\ \hline
        114 & Curaçao & CUW & North America \\ \hline
        115 & Dominican Republic & DOM & North America \\ \hline
        116 & Grenada & GRD & North America \\ \hline
        117 & Guadeloupe & GLP & North America \\ \hline
        118 & Haiti & HTI & North America \\ \hline
        119 & Jamaica & JAM & North America \\ \hline
        120 & Martinique & MTQ & North America \\ \hline
        121 & Puerto Rico & PRI & North America \\ \hline
        122 & Saint Lucia & LCA & North America \\ \hline
        123 & Saint Vincent and the Grenadines & VCT & North America \\ \hline
        124 & Trinidad and Tobago & TTO & North America \\ \hline
        125 & United States Virgin Islands & VIR & North America \\ \hline
        126 & Belize & BLZ & North America \\ \hline
        127 & Costa Rica & CRI & North America \\ \hline
        128 & El Salvador & SLV & North America \\ \hline
        129 & Guatemala & GTM & North America \\ \hline
        130 & Honduras & HND & North America \\ \hline
        131 & Mexico & MEX & North America \\ \hline
        132 & Nicaragua & NIC & North America \\ \hline
        133 & Panama & PAN & North America \\ \hline
        134 & Argentina & ARG & South America \\ \hline
        135 & Bolivia (Plurinational State of) & BOL & South America \\ \hline
        136 & Brazil & BRA & South America \\ \hline
        137 & Chile & CHL & South America \\ \hline
        138 & Colombia & COL & South America \\ \hline
        139 & Ecuador & ECU & South America \\ \hline
        140 & French Guiana & GUF & South America \\ \hline
        141 & Guyana & GUY & South America \\ \hline
        142 & Paraguay & PRY & South America \\ \hline
        143 & Peru & PER & South America \\ \hline
        144 & Suriname & SUR & South America \\ \hline
        145 & Uruguay & URY & South America \\ \hline
        146 & Venezuela (Bolivarian Republic of) & VEN & South America \\ \hline
        147 & Australia & AUS & Oceania \\ \hline
        148 & New Zealand & NZL & Oceania \\ \hline
        149 & Fiji & FJI & Oceania \\ \hline
        150 & New Caledonia & NCL & Oceania \\ \hline
        151 & Papua New Guinea & PNG & Oceania \\ \hline
        152 & Solomon Islands & SLB & Oceania \\ \hline
        153 & Vanuatu & VUT & Oceania \\ \hline
        154 & Guam & GUM & Oceania \\ \hline
        155 & Kiribati & KIR & Oceania \\ \hline
        156 & Micronesia (Fed. States of) & FSM & Oceania \\ \hline
        157 & French Polynesia & PYF & Oceania \\ \hline
        158 & Samoa & WSM & Oceania \\ \hline
        159 & Tonga & TON & Oceania \\ \hline
        160 & Belarus & BLR & Europe \\ \hline
        161 & Bulgaria & BGR & Europe \\ \hline
        162 & Czechia & CZE & Europe \\ \hline
        163 & Hungary & HUN & Europe \\ \hline
        164 & Poland & POL & Europe \\ \hline
        165 & Republic of Moldova & MDA & Europe \\ \hline
        166 & Romania & ROU & Europe \\ \hline
        167 & Russian Federation & RUS & Europe \\ \hline
        168 & Slovakia & SVK & Europe \\ \hline
        169 & Ukraine & UKR & Europe \\ \hline
        170 & Denmark & DNK & Europe \\ \hline
        171 & Estonia & EST & Europe \\ \hline
        172 & Finland & FIN & Europe \\ \hline
        173 & Iceland & ISL & Europe \\ \hline
        174 & Ireland & IRL & Europe \\ \hline
        175 & Latvia & LVA & Europe \\ \hline
        176 & Lithuania & LTU & Europe \\ \hline
        177 & Norway & NOR & Europe \\ \hline
        178 & Sweden & SWE & Europe \\ \hline
        179 & United Kingdom & GBR & Europe \\ \hline
        180 & Albania & ALB & Europe \\ \hline
        181 & Bosnia and Herzegovina & BIH & Europe \\ \hline
        182 & Croatia & HRV & Europe \\ \hline
        183 & Greece & GRC & Europe \\ \hline
        184 & Italy & ITA & Europe \\ \hline
        185 & Malta & MLT & Europe \\ \hline
        186 & Montenegro & MNE & Europe \\ \hline
        187 & North Macedonia & MKD & Europe \\ \hline
        188 & Portugal & PRT & Europe \\ \hline
        189 & Serbia & SRB & Europe \\ \hline
        190 & Slovenia & SVN & Europe \\ \hline
        191 & Spain & ESP & Europe \\ \hline
        192 & Austria & AUT & Europe \\ \hline
        193 & Belgium & BEL & Europe \\ \hline
        194 & France & FRA & Europe \\ \hline
        195 & Germany & DEU & Europe \\ \hline
        196 & Luxembourg & LUX & Europe \\ \hline
        197 & Netherlands & NLD & Europe \\ \hline
        198 & Switzerland & CHE & Europe \\ \hline
        199 & Canada & CAN & North America \\ \hline
        200 & United States of America & USA & North America \\ \hline
        ~ & ~ & ~ & ~ \\ \hline
\end{longtable}


\begin{thebibliography}{73}
	\providecommand{\natexlab}[1]{#1}
	\providecommand{\url}[1]{\texttt{#1}}
	\expandafter\ifx\csname urlstyle\endcsname\relax
	\providecommand{\doi}[1]{doi: #1}\else
	\providecommand{\doi}{doi: \begingroup \urlstyle{rm}\Url}\fi
	
	\bibitem[Aburto et~al.(2020)Aburto, Villavicencio, Basellini, Kj{\ae}rgaard,
	and Vaupel]{aburto2020dynamics}
	Jos{\'e}~Manuel Aburto, Francisco Villavicencio, Ugofilippo Basellini, S{\o}ren
	Kj{\ae}rgaard, and James~W Vaupel.
	\newblock Dynamics of life expectancy and life span equality.
	\newblock \emph{Proceedings of the National Academy of Sciences}, 117\penalty0
	(10):\penalty0 5250--5259, 2020.
	
	\bibitem[Afshar et~al.(2017)Afshar, Ho, Dilkina, Perros, Khalil, Xiong, and
	Sunderam]{afshar2017cp}
	Ardavan Afshar, Joyce~C Ho, Bistra Dilkina, Ioakeim Perros, Elias~B Khalil,
	Li~Xiong, and Vaidy Sunderam.
	\newblock Cp-ortho: An orthogonal tensor factorization framework for
	spatio-temporal data.
	\newblock In \emph{Proceedings of the 25th ACM SIGSPATIAL international
		conference on advances in geographic information systems}, pages 1--4, 2017.
	
	\bibitem[Aksoy and Poutvaara(2021)]{aksoy2021refugees}
	Cevat~Giray Aksoy and Panu Poutvaara.
	\newblock Refugees' and irregular migrants’ self-selection into {Europe}.
	\newblock \emph{Journal of Development Economics}, 152:\penalty0 102681, 2021.
	
	\bibitem[Anandkumar et~al.(2014)Anandkumar, Ge, Hsu, Kakade, and
	Telgarsky]{anandkumar2014tensor}
	Animashree Anandkumar, Rong Ge, Daniel Hsu, Sham~M Kakade, and Matus Telgarsky.
	\newblock Tensor decompositions for learning latent variable models.
	\newblock \emph{Journal of machine learning research}, 15:\penalty0 2773--2832,
	2014.
	
	\bibitem[Angel et~al.(2017)Angel, Vega, and L{\'o}pez-Ortega]{angel2017aging}
	Jacqueline~L Angel, William Vega, and Mariana L{\'o}pez-Ortega.
	\newblock Aging in mexico: Population trends and emerging issues.
	\newblock \emph{The Gerontologist}, 57\penalty0 (2):\penalty0 153--162, 2017.
	
	\bibitem[Bai and Lei(2020)]{bai2020new}
	Chen Bai and Xiaoyan Lei.
	\newblock New trends in population aging and challenges for {China}’s
	sustainable development.
	\newblock \emph{China Economic Journal}, 13\penalty0 (1):\penalty0 3--23, 2020.
	
	\bibitem[Beard et~al.(2012)Beard, Kalache, Delgado, and Hill]{beard2012ageing}
	John Beard, Alex Kalache, Mario Delgado, and Terry Hill.
	\newblock Ageing and urbanization.
	\newblock \emph{Global Population Ageing: Peril or Promise?}, pages 93--96,
	2012.
	
	\bibitem[Bijak and Bryant(2016)]{bijak2016bayesian}
	Jakub Bijak and John Bryant.
	\newblock Bayesian demography 250 years after bayes.
	\newblock \emph{Population studies}, 70\penalty0 (1):\penalty0 1--19, 2016.
	
	\bibitem[Bloom and Luca(2016)]{bloom2016global}
	David~E Bloom and Dara~Lee Luca.
	\newblock The global demography of aging: facts, explanations, future.
	\newblock In \emph{Handbook of the economics of population aging}, volume~1,
	pages 3--56. Elsevier, 2016.
	
	\bibitem[Bloom et~al.(2015)Bloom, Canning, and Lubet]{bloom2015global}
	David~E Bloom, David Canning, and Alyssa Lubet.
	\newblock Global population aging: Facts, challenges, solutions \&
	perspectives.
	\newblock \emph{Daedalus}, 144\penalty0 (2):\penalty0 80--92, 2015.
	
	\bibitem[Bro and Kiers(2003)]{bro2003new}
	Rasmus Bro and Henk~AL Kiers.
	\newblock A new efficient method for determining the number of components in
	parafac models.
	\newblock \emph{Journal of Chemometrics: A Journal of the Chemometrics
		Society}, 17\penalty0 (5):\penalty0 274--286, 2003.
	
	\bibitem[Chen et~al.(2019)Chen, Xu, Song, Wang, and He]{chen2019china}
	Rong Chen, Ping Xu, Peipei Song, Meifeng Wang, and Jiangjiang He.
	\newblock {China} has faster pace than {Japan} in population aging in next 25
	years.
	\newblock \emph{Bioscience trends}, 13\penalty0 (4):\penalty0 287--291, 2019.
	
	\bibitem[Chen et~al.(2021)Chen, Ying, Chen, Zhang, Lu, Fan, and
	Chen]{chen2021eigen}
	Xiaojie Chen, Na~Ying, Dean Chen, Yongwen Zhang, Bo~Lu, Jingfang Fan, and
	Xiaosong Chen.
	\newblock Eigen microstates and their evolution of global ozone at different
	geopotential heights.
	\newblock \emph{Chaos: An Interdisciplinary Journal of Nonlinear Science},
	31\penalty0 (7):\penalty0 071102, 2021.
	
	\bibitem[Cordesman(2018)]{cordesman2018kuwait}
	Anthony~H Cordesman.
	\newblock \emph{Kuwait: Recovery and security after the Gulf War}.
	\newblock Routledge, 2018.
	
	\bibitem[Danely(2015)]{danely2015aging}
	Jason Danely.
	\newblock \emph{Aging and loss: Mourning and maturity in contemporary Japan}.
	\newblock Rutgers University Press, 2015.
	
	\bibitem[D{\l}ugosz and Ra{\'z}niak(2014)]{dlugosz2014risk}
	Zbigniew D{\l}ugosz and Piotr Ra{\'z}niak.
	\newblock Risk of population aging in asia.
	\newblock \emph{Procedia-Social and Behavioral Sciences}, 120:\penalty0 36--45,
	2014.
	
	\bibitem[Fang et~al.(2015)Fang, Scheibye-Knudsen, Jahn, Li, Ling, Guo, Zhu,
	Preedy, Lu, Bohr, et~al.]{fang2015research}
	Evandro~Fei Fang, Morten Scheibye-Knudsen, Heiko~J Jahn, Juan Li, Li~Ling,
	Hongwei Guo, Xinqiang Zhu, Victor Preedy, Huiming Lu, Vilhelm~A Bohr, et~al.
	\newblock A research agenda for aging in {China} in the 21st century.
	\newblock \emph{Ageing research reviews}, 24:\penalty0 197--205, 2015.
	
	\bibitem[Fargues(2011)]{fargues2011immigration}
	Philippe Fargues.
	\newblock Immigration without inclusion: Non-nationals in nation-building in
	the gulf states.
	\newblock \emph{Asian and Pacific Migration Journal}, 20\penalty0
	(3-4):\penalty0 273--292, 2011.
	
	\bibitem[Garenne and Joseph(2002)]{garenne2002timing}
	Michel Garenne and Veronique Joseph.
	\newblock The timing of the fertility transition in sub-saharan africa.
	\newblock \emph{World Development}, 30\penalty0 (10):\penalty0 1835--1843,
	2002.
	
	\bibitem[Geddes et~al.(2020)Geddes, Einevoll, Acar, and
	Stasik]{geddes2020multi}
	Justen Geddes, Gaute~T Einevoll, Evrim Acar, and Alexander~J Stasik.
	\newblock Multi-linear population analysis (mlpa) of lfp data using tensor
	decompositions.
	\newblock \emph{Frontiers in Applied Mathematics and Statistics}, 6:\penalty0
	41, 2020.
	
	\bibitem[Gerland et~al.(2014)Gerland, Raftery, {\v{S}}ev{\v{c}}{\'\i}kov{\'a},
	Li, Gu, Spoorenberg, Alkema, Fosdick, Chunn, Lalic, et~al.]{gerland2014world}
	Patrick Gerland, Adrian~E Raftery, Hana {\v{S}}ev{\v{c}}{\'\i}kov{\'a}, Nan Li,
	Danan Gu, Thomas Spoorenberg, Leontine Alkema, Bailey~K Fosdick, Jennifer
	Chunn, Nevena Lalic, et~al.
	\newblock World population stabilization unlikely this century.
	\newblock \emph{Science}, 346\penalty0 (6206):\penalty0 234--237, 2014.
	
	\bibitem[{Global Media Insight}(2023)]{ARE@Misc}
	{Global Media Insight}.
	\newblock United arab emirates population statistics 2023.
	\newblock
	\url{https://www.globalmediainsight.com/blog/uae-population-statistics/},
	2023.
	\newblock Accessed: 2023-3-28.
	
	\bibitem[Goh et~al.(2020)Goh, McNown, et~al.]{goh2020macroeconomic}
	Soo~Khoon Goh, Robert McNown, et~al.
	\newblock Macroeconomic implications of population aging: Evidence from japan.
	\newblock \emph{Journal of Asian Economics}, 68:\penalty0 101198, 2020.
	
	\bibitem[Gu et~al.(2021)Gu, Andreev, and Dupre]{gu2021major}
	Danan Gu, Kirill Andreev, and Matthew~E Dupre.
	\newblock Major trends in population growth around the world.
	\newblock \emph{{China} CDC weekly}, 3\penalty0 (28):\penalty0 604, 2021.
	
	\bibitem[Harper(2014)]{harper2014economic}
	Sarah Harper.
	\newblock Economic and social implications of aging societies.
	\newblock \emph{Science}, 346\penalty0 (6209):\penalty0 587--591, 2014.
	
	\bibitem[He et~al.(2016)He, Goodkind, Kowal, et~al.]{he2016aging}
	Wan He, Daniel Goodkind, Paul~R Kowal, et~al.
	\newblock An aging world: 2015.
	\newblock
	\url{https://www.researchgate.net/profile/Paul-Kowal/publication/299528572_An_Aging_World_2015/links/56fd4be108ae17c8efaa1132/An-Aging-World-2015.pdf},
	2016.
	\newblock Accessed: 2023-3-27.
	
	\bibitem[Hidalgo(2021)]{2021Economic}
	C{\'e}sar~A Hidalgo.
	\newblock Economic complexity theory and applications.
	\newblock \emph{Nature Reviews Physics}, 3\penalty0 (2):\penalty0 92--113,
	2021.
	
	\bibitem[Hidalgo et~al.(2007)Hidalgo, Klinger, Barab{\'a}si, and
	Hausmann]{Hidalgo2007The}
	C{\'e}sar~A Hidalgo, Bailey Klinger, A-L Barab{\'a}si, and Ricardo Hausmann.
	\newblock The product space conditions the development of nations.
	\newblock \emph{Science}, 317\penalty0 (5837):\penalty0 482--487, 2007.
	
	\bibitem[Hu(2018)]{hu2018universality}
	Chin-Kun Hu.
	\newblock Universality and scaling in human and social systems.
	\newblock In \emph{Journal of Physics: Conference Series}, volume 1113, page
	012002. IOP Publishing, 2018.
	
	\bibitem[Hu et~al.(2019)Hu, Liu, Liu, Chen, and Chen]{hu2019condensation}
	GaoKe Hu, Teng Liu, MaoXin Liu, Wei Chen, and XiaoSong Chen.
	\newblock Condensation of eigen microstate in statistical ensemble and phase
	transition.
	\newblock \emph{Science China Physics, Mechanics \& Astronomy}, 62\penalty0
	(9):\penalty0 1--8, 2019.
	
	\bibitem[Huang et~al.(2020)Huang, Maller, and Ning]{huang2020modelling}
	Fei Huang, Ross Maller, and Xu~Ning.
	\newblock Modelling life tables with advanced ages: An extreme value theory
	approach.
	\newblock \emph{Insurance: Mathematics and Economics}, 93:\penalty0 95--115,
	2020.
	
	\bibitem[Huang et~al.(2021)Huang, Li, and Chen]{huang2021longcuts}
	Siyu Huang, Xiaomeng Li, and Qinghua Chen.
	\newblock Longcuts in the global migration network.
	\newblock \emph{EPL (Europhysics Letters)}, 134\penalty0 (4):\penalty0 48002,
	2021.
	
	\bibitem[Impagliazzo(2012)]{impagliazzo2012deterministic}
	John Impagliazzo.
	\newblock \emph{Deterministic aspects of mathematical demography: An
		investigation of the stable theory of population including an analysis of the
		population statistics of Denmark}, volume~13.
	\newblock Springer Science \& Business Media, 2012.
	
	\bibitem[Jones(2020)]{jones2020problem}
	Katie Jones.
	\newblock The problem of an aging global population, shown by country.
	\newblock \emph{Geography Bulletin}, 52\penalty0 (1):\penalty0 21--23, 2020.
	
	\bibitem[Kadu{\v{s}}i{\'c} et~al.(2016)Kadu{\v{s}}i{\'c}, Sulji{\'c}, and
	Smaji{\'c}]{kaduvsic2016demographic}
	Alma Kadu{\v{s}}i{\'c}, Alija Sulji{\'c}, and Nedima Smaji{\'c}.
	\newblock The demographic ageing of population in bosnia and herzegovina:
	causes and consequences.
	\newblock \emph{Revija za geografijo}, 11\penalty0 (1):\penalty0 41--52, 2016.
	
	\bibitem[Kashnitsky and Aburto(2020)]{kashnitsky2020covid}
	Ilya Kashnitsky and Jos{\'e}~Manuel Aburto.
	\newblock Covid-19 in unequally ageing european regions.
	\newblock \emph{World development}, 136:\penalty0 105170, 2020.
	
	\bibitem[Khan et~al.(2022)Khan, Hou, Zakari, Irfan, and Ahmad]{khan2022links}
	Irfan Khan, Fujun Hou, Abdulrasheed Zakari, Muhammad Irfan, and Munir Ahmad.
	\newblock Links among energy intensity, non-linear financial development, and
	environmental sustainability: New evidence from asia pacific economic
	cooperation countries.
	\newblock \emph{Journal of Cleaner Production}, 330:\penalty0 129747, 2022.
	
	\bibitem[Kolda and Bader(2009)]{kolda2009tensor}
	Tamara~G Kolda and Brett~W Bader.
	\newblock Tensor decompositions and applications.
	\newblock \emph{SIAM review}, 51\penalty0 (3):\penalty0 455--500, 2009.
	
	\bibitem[Lee and Mason(2011)]{lee2011population}
	Ronald~Demos Lee and Andrew Mason.
	\newblock \emph{Population aging and the generational economy: A global
		perspective}.
	\newblock Edward Elgar Publishing, 2011.
	
	\bibitem[Li and Chen(2016)]{li2016critical}
	Xiao-Teng Li and Xiao-Song Chen.
	\newblock Critical behaviors and finite-size scaling of principal fluctuation
	modes in complex systems.
	\newblock \emph{Communications in Theoretical Physics}, 66\penalty0
	(3):\penalty0 355, 2016.
	
	\bibitem[Li et~al.(2016)Li, Xu, Chen, Chen, Zhang, and
	Di]{li2016characterizing}
	Xiaomeng Li, Hongzhong Xu, Jiawei Chen, Qinghua Chen, Jiang Zhang, and Zengru
	Di.
	\newblock Characterizing the international migration barriers with a
	probabilistic multilateral migration model.
	\newblock \emph{Scientific reports}, 6\penalty0 (1):\penalty0 1--14, 2016.
	
	\bibitem[Li et~al.(2021)Li, Xue, Sun, Fan, Li, Liu, Han, Di, and
	Chen]{li2021discontinuous}
	Xu~Li, Tingting Xue, Yu~Sun, Jingfang Fan, Hui Li, Maoxin Liu, Zhangang Han,
	Zengru Di, and Xiaosong Chen.
	\newblock Discontinuous and continuous transitions of collective behaviors in
	living systems.
	\newblock \emph{Chinese Physics B}, 30\penalty0 (12):\penalty0 128703, 2021.
	
	\bibitem[Liu and Raftery(2020)]{liu2020education}
	Daphne~H Liu and Adrian~E Raftery.
	\newblock How do education and family planning accelerate fertility decline?
	\newblock \emph{Population and development review}, 46\penalty0 (3):\penalty0
	409--441, 2020.
	
	\bibitem[Liu(2010)]{liu2010china}
	Lee Liu.
	\newblock {China}'s population trends and their implications for fertility
	policy.
	\newblock \emph{Asian Population Studies}, 6\penalty0 (3):\penalty0 289--305,
	2010.
	
	\bibitem[Liu et~al.(2022)Liu, Hu, Dong, Fan, Liu, and
	Chen]{liu2022renormalization}
	Teng Liu, Gao-Ke Hu, Jia-Qi Dong, Jing-Fang Fan, Mao-Xin Liu, and Xiao-Song
	Chen.
	\newblock Renormalization group theory of eigen microstates.
	\newblock \emph{Chinese Physics Letters}, 39\penalty0 (8):\penalty0 080503,
	2022.
	
	\bibitem[Mback{\'e}(2017)]{mbacke2017persistence}
	Cheikh Mback{\'e}.
	\newblock The persistence of high fertility in sub-saharan africa: a comment.
	\newblock \emph{Population and Development Review}, 43:\penalty0 330--337,
	2017.
	
	\bibitem[Miranda et~al.(2016)Miranda, Mendes, and Silva]{miranda2016population}
	Gabriella Morais~Duarte Miranda, Antonio da Cruz~Gouveia Mendes, and Ana Lucia
	Andrade~da Silva.
	\newblock Population aging in brazil: current and future social challenges and
	consequences.
	\newblock \emph{Revista Brasileira de Geriatria e Gerontologia}, 19:\penalty0
	507--519, 2016.
	
	\bibitem[Mizoguchi(2010)]{mizoguchi2010consequences}
	Nobuko Mizoguchi.
	\newblock \emph{The consequences of the Vietnam war on the Vietnamese
		population}.
	\newblock University of California, Berkeley, 2010.
	
	\bibitem[Mountford and Rapoport(2016)]{2016Migration}
	A.~Mountford and H.~Rapoport.
	\newblock Migration policy, african population growth and global inequality.
	\newblock \emph{World Economy}, 39\penalty0 (4):\penalty0 543--556, 2016.
	
	\bibitem[Natali(2010)]{natali2010kurdish}
	Denise Natali.
	\newblock \emph{The Kurdish quasi-state: Development and dependency in
		post-Gulf War Iraq}.
	\newblock Syracuse University Press, 2010.
	
	\bibitem[Pavl{\'\i}k(2000)]{pavlik2000demography}
	Z~Pavl{\'\i}k.
	\newblock What is demography.
	\newblock \emph{Position of demography among other disciplines. Charles
		University, Faculty of Science, Prague}, pages 9--18, 2000.
	
	\bibitem[Permanyer and Scholl(2019)]{permanyer2019global}
	I{\~n}aki Permanyer and Nathalie Scholl.
	\newblock Global trends in lifespan inequality: 1950-2015.
	\newblock \emph{PloS one}, 14\penalty0 (5):\penalty0 e0215742, 2019.
	
	\bibitem[Pezzulo et~al.(2021)Pezzulo, Nilsen, Carioli, Tejedor~Garavito,
	Hanspal, Hilber, James, Ruktanonchai, Alegana, Sorichetta,
	et~al.]{pezzulo2021geography}
	Carla Pezzulo, Kristine Nilsen, Alessandra Carioli, Natalia Tejedor~Garavito,
	Sophie~E Hanspal, Theodor Hilber, William~HM James, Corrine~Warren
	Ruktanonchai, VA~Alegana, Alessandro Sorichetta, et~al.
	\newblock The geography of fertility rates in low and middle-income countries:
	a subnational analysis of cross-sectional surveys from 70 countries, 2010-16.
	\newblock \emph{The Lancet Global Health}, 9\penalty0 (6):\penalty0 E802--E812,
	2021.
	
	\bibitem[Raducha and Gubiec(2017)]{2017Coevolving}
	T.~Raducha and T.~Gubiec.
	\newblock Coevolving complex networks in the model of social interactions.
	\newblock \emph{Physica A: Statistical Mechanics and its Applications}, 2017.
	
	\bibitem[Ritchie and Roser(2019{\natexlab{a}})]{owidagestructure}
	Hannah Ritchie and Max Roser.
	\newblock Age structure.
	\newblock \emph{Our World in Data}, 2019{\natexlab{a}}.
	\newblock https://ourworldindata.org/age-structure.
	
	\bibitem[Ritchie and Roser(2019{\natexlab{b}})]{ritchie2019age}
	Hannah Ritchie and Max Roser.
	\newblock Age structure.
	\newblock \emph{Our World in Data}, 2019{\natexlab{b}}.
	\newblock https://ourworldindata.org/age-structure.
	
	\bibitem[Sanders(2021)]{sanders2021125}
	S~Sanders.
	\newblock 125 questions: Exploration and discovery.
	\newblock \emph{Science/AAAS Custom Publishing Office: Washington, DC, USA},
	2021.
	
	\bibitem[Scheidel(2001)]{scheidel2001debating}
	Walter Scheidel.
	\newblock \emph{Debating Roman Demography}, volume 211.
	\newblock Brill, 2001.
	
	\bibitem[Sheets and Gallagher(2013)]{sheets2013aging}
	Debra~J Sheets and Elaine~M Gallagher.
	\newblock Aging in canada: state of the art and science.
	\newblock \emph{The Gerontologist}, 53\penalty0 (1):\penalty0 1--8, 2013.
	
	\bibitem[Strang et~al.(1993)Strang, Strang, Strang, and
	Strang]{strang1993introduction}
	Gilbert Strang, Gilbert Strang, Gilbert Strang, and Gilbert Strang.
	\newblock \emph{Introduction to linear algebra}, volume~3.
	\newblock Wellesley-Cambridge Press Wellesley, MA, 1993.
	
	\bibitem[Strulik and Vollmer(2015)]{strulik2015fertility}
	Holger Strulik and Sebastian Vollmer.
	\newblock The fertility transition around the world.
	\newblock \emph{Journal of Population Economics}, 28\penalty0 (1):\penalty0
	31--44, 2015.
	
	\bibitem[Sun et~al.(2021)Sun, Hu, Zhang, Lu, Lu, Fan, Li, Deng, and
	Chen]{sun2021eigen}
	Yu~Sun, Gaoke Hu, Yongwen Zhang, Bo~Lu, Zhenghui Lu, Jingfang Fan, Xiaoteng Li,
	Qimin Deng, and Xiaosong Chen.
	\newblock Eigen microstates and their evolutions in complex systems.
	\newblock \emph{Communications in Theoretical Physics}, 73\penalty0
	(6):\penalty0 065603, 2021.
	
	\bibitem[Tacchella et~al.(2018)Tacchella, Mazzilli, and Pietronero]{2018nature}
	Andrea Tacchella, Dario Mazzilli, and Luciano Pietronero.
	\newblock A dynamical systems approach to gross domestic product forecasting.
	\newblock \emph{Nature Physics}, 14\penalty0 (8):\penalty0 861--865, 2018.
	
	\bibitem[Tey et~al.(2016)Tey, Siraj, Kamaruzzaman, Chin, Tan, Sinnappan, and
	M{\"u}ller]{tey2016aging}
	Nai~Peng Tey, Saedah~Binti Siraj, Shahrul Bahyah~Binti Kamaruzzaman, Ai~Vyrn
	Chin, Maw~Pin Tan, Glaret~Shirley Sinnappan, and Andre~Matthias M{\"u}ller.
	\newblock Aging in multi-ethnic malaysia.
	\newblock \emph{The Gerontologist}, 56\penalty0 (4):\penalty0 603--609, 2016.
	
	\bibitem[{United Nations}(2023)]{Niger@Misc}
	{United Nations}.
	\newblock Population ages 0-14 (\% of total population) - niger.
	\newblock
	\url{https://data.worldbank.org/indicator/SP.POP.0014.TO.ZS?locations=NE},
	2023.
	\newblock Accessed: 2023-3-28.
	
	\bibitem[United~Nations and Social~Affairs(2022)]{united2022world}
	Department of~Economic United~Nations and Population~Division Social~Affairs.
	\newblock World population prospects 2022: summary of results, 2022.
	
	\bibitem[Wang et~al.(2017)Wang, Abajobir, Abate, Abbafati, Abbas, Abd-Allah,
	Abera, Abraha, Abu-Raddad, Abu-Rmeileh, et~al.]{wang2017global}
	Haidong Wang, Amanuel~Alemu Abajobir, Kalkidan~Hassen Abate, Cristiana
	Abbafati, Kaja~M Abbas, Foad Abd-Allah, Semaw~Ferede Abera, Haftom~Niguse
	Abraha, Laith~J Abu-Raddad, Niveen~ME Abu-Rmeileh, et~al.
	\newblock Global, regional, and national under-5 mortality, adult mortality,
	age-specific mortality, and life expectancy, 1970--2016: a systematic
	analysis for the global burden of disease study 2016.
	\newblock \emph{The Lancet}, 390\penalty0 (10100):\penalty0 1084--1150, 2017.
	
	\bibitem[Wang and Sun(2016)]{wang2016role}
	Qingfeng Wang and Xu~Sun.
	\newblock The role of socio-political and economic factors in fertility
	decline: a cross-country analysis.
	\newblock \emph{World Development}, 87:\penalty0 360--370, 2016.
	
	\bibitem[Woods(2007)]{woods2007ancient}
	Robert Woods.
	\newblock Ancient and early modern mortality: experience and understanding 1.
	\newblock \emph{The Economic History Review}, 60\penalty0 (2):\penalty0
	373--399, 2007.
	
	\bibitem[{World Population Review}(2023{\natexlab{a}})]{Oman@Misc}
	{World Population Review}.
	\newblock Oman population 2023 (live).
	\newblock \url{https://worldpopulationreview.com/countries/oman-population},
	2023{\natexlab{a}}.
	\newblock Accessed: 2023-3-28.
	
	\bibitem[{World Population Review}(2023{\natexlab{b}})]{Qatar@Misc}
	{World Population Review}.
	\newblock Qatar population 2023 (live).
	\newblock \url{https://worldpopulationreview.com/countries/qatar-population},
	2023{\natexlab{b}}.
	\newblock Accessed: 2023-3-28.
	
	\bibitem[Zeng and Ng(2020)]{zeng2020decompositions}
	Chao Zeng and Michael~K Ng.
	\newblock Decompositions of third-order tensors: Hosvd, t-svd, and beyond.
	\newblock \emph{Numerical Linear Algebra with Applications}, 27\penalty0
	(3):\penalty0 e2290, 2020.
	
	\bibitem[Zhao(2010)]{zhao2010historical}
	Zhongwei Zhao.
	\newblock Historical demography.
	\newblock \emph{Demography}, 1:\penalty0 38--64, 2010.
	
\end{thebibliography}
\end{document}